\documentclass[12pt]{article}

\usepackage{amsmath,amssymb,array,amsfonts}
\usepackage[dvipdfmx]{graphicx}
\usepackage{comment,color}

\newcommand{\bs}{\boldsymbol}

\allowdisplaybreaks

\def\e {{\rm e}}	
\def\s {{\rm s}}

\def\H {{\rm H}}
\def\SM{{\rm SM}}
\def\T {{\rm T}}
\def\tr{{\rm Tr}}
\def\tt{\times10}

\def\gm{\gamma^{\mu}}		
			
\def\bls{\baselineskip}
\def\dis{\displaystyle}
\def\lra{\longrightarrow}

\def\sig{i\sigma^2}

\def\dy{\int_{-\pi R}^{\pi R}\!\!\!\!\!\!\!\!{\rm d}y\,}%

\def\nt{\notag}                                

\makeatletter
    
    \@addtoreset{equation}{section}
\makeatother

\begin{document}
\setlength{\baselineskip}{18pt}
\begin{titlepage}

\begin{flushright}
KOBE-TH-11-04
\end{flushright}
\vspace{1.0cm}
\begin{center}
{\Large\bf $\bs{D^0}$\,--\,$\bs{\bar D^0}$ Mixing in Gauge-Higgs Unification} 
\end{center}
\vspace{25mm}

\begin{center}
{\large
Yuki Adachi, 
%
Nobuaki Kurahashi$^*$, 
%
C. S. Lim$^*$
%
%
and Nobuhito Maru$^{**}$
}
\end{center}
\vspace{1cm}
\centerline{{\it
Department of Sciences, Matsue College of Technology,
Matsue 690-8518, Japan}}

\centerline{{\it
$^*$Department of Physics, Kobe University,
Kobe 657-8501, Japan}}

\centerline{{\it
$^{**}$Department of Physics, Chuo University, 
Tokyo 112-8551, Japan\footnote{
Present address:  Department of Physics,
and Research and Education Center for Natural Sciences, 
Keio University, Hiyoshi, Yokohama, 223-8521 Japan.}
}}
%
%
\vspace{2cm}
\centerline{\large\bf Abstract}
\vspace{0.5cm}

We discuss flavor mixing and resulting Flavor Changing Neutral Current (FCNC) in the $SU(3) \otimes SU(3)_\text{color}$ gauge-Higgs unification. As the FCNC process we calculate the rate of $D^0-\bar{D}^0$ mixing due to the exchange of non-zero Kaluza-Klein gluons at the tree level. Flavor mixing is argued to be realized by the fact that the bulk mass term and brane localized mass term is not diagonalized simultaneously unless bulk masses are degenerate. 
It is shown that automatic suppression 
mechanism is operative for the FCNC processes of light quarks. We therefore obtain a lower bound on the compactification scale of order ${\cal O}({\rm TeV})$ by comparing our prediction on the mass difference of neutral $D$ meson with the recent experimental data, which is much milder than what we naively expect assuming only the decoupling of non-zero Kaluza-Klein gluons. 

\end{titlepage}




\newpage

\section{Introduction}

The origin of electroweak gauge symmetry breaking 
is still mysterious in particle physics. Though in the Standard Model (SM), 
Higgs boson is assumed to play a role for the symmetry breaking, it seems to have various 
theoretical problems such as the hierarchy problem and the presence of many theoretically unpredicted 
arbitrary coupling constants in its interactions.  

Gauge-Higgs unification (GHU) \cite{GH}
is one of the attractive scenarios to go beyond the standard model. 
It provides a possible solution to the hierarchy problem without invoking supersymmetry, 
also shedding some light on the long standing arbitrariness problem of Higgs interactions. 
In this scenario, Higgs boson in the SM is identified
with the extra spatial components of the higher dimensional gauge fields. 
Remarkable feature is that the quantum correction to Higgs mass is UV-finite and calculable without invoking supersymmetry 
regardless of the non-renormalizability of higher dimensional gauge theory. 
This feature is guaranteed by the higher dimensional gauge invariance and has opened up a new avenue to the solution to the hierarchy problem \cite{HIL}. 
The finiteness of the Higgs mass has been studied and verified in various models 
and types of compactification at one-loop level\footnote{For the case of gravity-gauge-Higgs unification, see \cite{HLM}.}
\cite{ABQ} and even at the two loop level \cite{MY}. The fact that the Higgs boson is a part of gauge 
fields implies that Higgs interactions are restricted by gauge principle and may provide a possibility to 
solve the arbitrariness problem of Higgs interactions.

{}From such point of view, it seems that for the GHU to be phenomenologically viable, 
the following issues have special importance. 

\noindent (1)
Are there any characteristic and generic predictions
on the observables, which are subject to precision tests ?  

\noindent (2)
How are the flavor structure of fermion masses and flavor mixings realized in the Yukawa couplings starting from 
higher dimensional gauge interaction ?  

\noindent (3)
In view of the fact that Higgs interactions are basically gauge interactions
with real gauge coupling constants, 
how is CP violated ?  


As for the issue (1), 
it will be desirable to find finite (UV-insensitive) and calculable observables, in spite of the fact 
that the theory is non-renormalizable and observables are very UV-sensitive in general. 
Works on the oblique electroweak parameters and fermion anomalous magnetic moment from such a viewpoint 
already have been done in the literature \cite{LM}-\cite{ALM1}. 
The issue (3) has been addressed in our previous papers \cite{ALM3}, \cite{LMN}, 
where CP violation is claimed to be achieved ``spontaneously" either by the VEV of the Higgs field 
or by the ``complex structure" of the compactified extra space.  

In this paper, we focus on the remaining issue (2) concerning the flavor physics in the GHU scenario. 
It is highly non-trivial problem to explain the variety of fermion masses 
and flavor mixings in this scenario, 
since the gauge interactions should be universal for all matter fields,
while the flavor symmetry has to be broken eventually in order to distinguish each flavor and  
to realize their mixings.   
In our previous paper \cite{AKLM}, we 
addressed this issue and have clarified the mechanism to generate the flavor mixings 
by the interplay between bulk masses and the brane localized masses. As a remarkable property of higher dimensional 
gauge theories gauge invariant bulk mass terms are allowed in the form of sign functions of the extra space coordinate $y$. 
In the SM $SU(2)$ left-handed doublet of fermions couples to both of up-type and down-type right-handed fermions simultaneously. In the GHU, however, the up- and down-type right-handed fermions belong to different representations of gauge group, in general. Thus they couple to two independent left-handed doublets to form Yukawa couplings and therefore 
it is needed to introduce brane-localized fermions such that they form brane-localized mass terms with some linear combinations of the doublets in order to eliminate the redundant degrees of freedom. 

Important point is that such introduced two types of mass terms generically may be flavor non-diagonal without contradicting with gauge invariance. This property then leads to the flavor mixing in the up- and down-types of Yukawa couplings \cite{BN}. We may start with the base where the bulk mass terms are diagonalized, since the bulk mass terms are written in the form of hermitian matrix, which may be diagonalized by suitable unitary transformations, keeping the kinetic and gauge interaction terms of fermions invariant \cite{AKLM}. Even in this base, however, the brane-localized mass terms still have off-diagonal elements in the flavor base in general. Namely, the fact that in general two types of fermion mass terms cannot be diagonalized simultaneously leads to physical flavor mixing. This is why we stress that the interplay between these two types of mass terms is crucial.

At first thought, one might think that only the brane localized mass terms are enough 
to generate the flavor mixings since they can be put by hand. 
However, it is not the case. We have shown that the flavor mixings disappear in the limit of universal bulk masses 
where the hierarchy of fermion masses is absent \cite{AKLM}. The reason is in this limit the bulk mass terms remain flavor-diagonal for arbitrary unitary transformation of each representation of bulk fermions. By use of this degree 
of freedom the Yukawa couplings are readily made diagonal. This is a remarkable feature of the GHU scenario, 
which is not shared by, e.g., the universal extra dimension 
where the flavor mixing may be caused by Yukawa couplings in the bulk
just as in the standard model.  

Once the flavor mixings are realized, 
it will be important to discuss flavor changing neutral current (FCNC) processes, 
which have been playing a crucial role for checking the viability of various new physics models, as is seen in the case of SUSY model. 
This issue was first discussed in \cite{DPQ} in the context of extra dimensions. 
Since our model reduces to the SM at low energies, there is no FCNC processes at the tree level with respect to the zero mode fields. 
It, however, turns out that the exchange of non-zero Kaluza-Klein (KK) modes of gauge bosons causes FCNC at the tree level, 
though the rates of FCNC are suppressed by the inverse powers of the compactification scale (``decoupling") \cite{AKLM}. 
The reason is the following.  

As a genuine feature of the higher dimensional gauge theories with orbifold compactification, the gauge invariant bulk mass terms for fermions, generically written as $M \epsilon (y) \bar{\psi} \psi$ with $\epsilon (y)$ 
being the sign function of extra space coordinate $y$, are allowed. 
The bulk mass $M$ may be different depending on each generation
and can be an important new source of the flavor violation. 
The presence of the mass terms causes the localization of Weyl fermions
in two different fixed points of the orbifold depending on their chiralities
and the Yukawa coupling obtained by the overlap integral over $y$
of the mode functions of Weyl fermions with different chiralities
is suppressed by a factor $2\pi RM \e^{-\pi RM}
\ (R: \text{the size of the  extra space})$, 
which is otherwise just gauge coupling $g$ and universal for all flavors.
Thus in GHU scenario, fermion masses are all equal and of weak scale $M_{W}$ to start with 
and the observed hierarchical small fermion masses can be achieved without fine tuning thanks to the exponential suppression factor $\e^{-\pi RM}$. 
On the other hand, this means that the criteria by Glashow-Weinberg \cite{GW} in 4D space-time is not enough to ensure natural flavor conservation. Namely, the gauge couplings of non-zero KK modes of gauge boson, whose mode functions are $y$-dependent, are no longer universal even for Weyl fermions with definite chirality and the same quantum numbers, since the 
overlap integral of mode function of fermion and KK gauge boson depends on the bulk mass $M$. Thus once we move to the base of mass-eigenstates FCNC appears at the tree level. 

In the previous paper, as a typical process of FCNC, we calculated the $K^0-\bar{K}^0$ mixing amplitude at the tree level 
via non-zero KK gluon exchange and obtained the lower bounds for the compactification scale 
as the predictions of our model \cite{AKLM}.\footnote{Constraints from $K^0-\bar{K}^0$ mixing have been discussed in \cite{KT} 
for a similar model as ours although it is not the gauge-Higgs unification. Similar suppression mechanism of FCNC for light quarks to the one discussed in this paper has been pointed out to be operative in this reference.} 
Interestingly, the obtained lower bound of ${\cal O}(10)$ TeV was much milder than we naively expect assuming that the amplitude is simply suppressed by the inverse powers of the compactification scale, say ${\cal O}(10^{3})$ TeV . 
We pointed out the presence of suppression mechanism of the FCNC process, operative for light fermions in the GHU model. As was mentioned above, 
fermion masses much smaller than $M_W$ are realized by the localizations of fermions. Larger the bulk mass $M$, the localization of fermion is steeper and therefore for the fermions the mode functions of KK gluons seem to be almost constant. Thus for light fermions the gauge couplings of KK gluons become almost universal, just as in the case of the zero-mode sector. 

In the analysis, our mechanisms of the flavor mixing and the suppression of FCNC were applied to the down-type quark sector, 
but the mechanisms should be also applicable to the up-type quark sector. In this paper, we turn to the $D^0-\bar{D}^0$ mixing, which is caused by the mixing between up and charm quarks. The $D^0-\bar{D}^0$ mixing is not only the typical FCNC process in up-type quark sector, but also plays special role in exploring physics beyond the SM. Namely, in the SM the  $\Delta C = 2$ FCNC process is realized through ``box diagram" where internal quarks are of down-type, though 
in addition to such ``short distance" contribution poorly known ``long distance" contribution due to non-perturbative QCD effects are claimed to be important. The mass-squared differences of down-type quarks are much smaller than those of up-type quarks. Thus the expected contribution to the mass difference of neutral $D$ meson $\Delta M_{D}({\rm SD})$ due to $D^0-\bar{D}^0$ mixing is expected to be small in the SM: 
\begin{equation}
   x_{D}({\rm SM})
 = \frac{\Delta M_{D}({\rm SM})}{\Gamma_D} \lesssim 10^{-3},   
\end{equation}
where $\Gamma_{D}$ is the decay width of neutral $D$ meson. Hence if the $D^0-\bar{D}^0$ mixing and/or associated CP violating observable with relatively large rates are found it suggests the presence of some new physics. As the matter of fact, 
recently impressive progress has been made by BABAR and Belle in the measurement \cite{HFAG}: 
\begin{equation}
x_D({\rm exp}) 
 = (1.00 \pm 0.25) \times 10^{-2}.  
\end{equation}

We will calculate the dominant contribution to the process at the tree level 
by the exchange of non-zero KK gluons. Comparing the obtained finite contribution to the mixing with the allowed range for the new physics contribution derived from the experimental data, we put the lower bound on the compactification scale.   
It will be also discussed how the extent of the suppression of FCNC process is different depending on the type of contributing effective 4-Fermi operators, i.e. the operators made by the product of currents with the same chirality (LL and RR type) and different chiralities (LR type).

This paper is organized as follows.
After introducing our model in the next section, we summarize in section 3 how the flavor mixing is realized 
in the context of the gauge-Higgs unification, which was clarified and described in detail in our previous paper \cite{AKLM}. 
In section 4, as an application of the flavor mixing discussed in section 3, 
we calculate the mass difference of neutral $D$-mesons
caused by the $D^0-\bar{D}^0$ mixing via non-zero KK gluon exchange at the tree level.  
We also obtain the lower bound for the compactification scale 
by comparing the obtained result with the experimental data. 
The origin of the suppression mechanism of FCNC process is discussed in section 5, emphasizing the importance of the localization of quark fields and the fact that FCNC is 
controlled by the non-degeneracy of bulk masses, which is specific to the gauge-Higgs unification. Also discussed is the origin of the different extent of the suppression depending on the chirality of the relevant 4-Fermi operator. 
Section 6 is devoted to our conclusions.  
The results of more careful and thorough study of $K^0-\bar{K}^0$ mixing taking into account the contributions of the operators of LL and RR types, which was not carried out in our previous paper \cite{AKLM}, are briefly given in Appendix A.


\section{The Model}

Although the model we consider in this paper is the same as the one taken in \cite{AKLM}, 
we briefly describe the model for completeness. 
The model taken in this paper is a five dimensional (5D) $SU(3) \otimes SU(3)_\text{color}$ GHU model 
compactified on an orbifold $S^1/Z_2$ with a radius $R$ of $S^1$. 
As matter fields, 
we introduce 
$n$ generations of bulk fermion in the ${\bf 3}$ and the $\bar{{\bf 6}}$ 
dimensional representations of $SU(3)$ gauge group denoted by a column vector and a $3 \times 3$ matrix,  
$\psi^i({\bf 3})$ 
and
$\psi^i(\bar {\bf 6}) \ (i = 1, \ldots, n)$ \cite{BN}.

The bulk Lagrangian is given by 
\begin{align}
	\mathcal{L}
 =&	-\!\frac12\tr\big(F_{MN}F^{MN}\big)
	-\frac12\tr\big(G_{MN}G^{MN}\big)
	\notag\\*
  &	+\bar\psi^i({\bf 3})\big\{i\!\not\!\!D_3 -M^i\epsilon(y)\big\}\psi^i({\bf 3})
	+\frac12
	 \tr\Big[
	 \bar\psi^i({\bar{\bf 6}})\big\{i\!\not \!\!D_6 -M^i\epsilon(y)\big\}\psi^i({\bar{\bf 6}})
	 \Big]
\end{align}
where
\begin{align}
	F_{MN}
 &=	\partial_MA_N-\partial_NA_M -ig\big[A_M,A_N\big],\\
	G_{MN}
 &=	\partial_MG_N-\partial_NG_M -ig_\s\big[G_M,G_N\big],\\
	\not\!\!D_3 \psi^i({\bf 3}) 
 &= \Gamma^M(\partial_M-igA_M-ig_\s G_M)\psi^i({\bf 3}),\\
 	\not\!\!D_6 \psi^i({\bar{\bf 6}}) 
 &= \Gamma^M
	\Big[
	\partial_M\psi^i({\bar{\bf 6}})
	+ig\big\{A_M^{\ast}\psi^i({\bar{\bf 6}})+\psi^i({\bar{\bf 6}})(A_{M})^\dagger\big\}
	-ig_\s G_M\psi^i({\bar{\bf 6}})
	\Big],
\end{align}
with $G_M$ being understood to act on the color index, not explicitly written here. 
The gauge fields $A_M$ and $G_M$ are written in a matrix form,
e.g. $A_M = A_M^a \frac{\lambda^a}2$ in terms of Gell-Mann matrices $\lambda^{a}$. 
$M,N=0,1,2,3,5$ and the five dimensional gamma matrices
are given by $\Gamma^M=(\gamma^{\mu},i\gamma^{5})$ ($\mu=0,1,2,3$). 
$g$ and $g_\s$ are 5D gauge coupling constants
of $SU(3)$ and $SU(3)_\text{color}$, respectively. 
$M^i$ are generation dependent bulk mass parameters of the fermions
accompanied by the sign function $\epsilon (y)$. As was discussed in the introduction, here we take the base where 
the bulk mass term is flavor-diagonal.    

The periodic boundary condition is imposed along $S^1$ and $Z_2$ parity
assignments are taken for gauge fields as
\begin{align}
	A_{\mu}
 &= \left[
	\begin{array}{ccc}
	 (+,+)&(+,+)&(-,-)\\
	 (+,+)&(+,+)&(-,-)\\
	 (-,-)&(-,-)&(+,+)
	\end{array}
	\right],
 ~~
	A_y
 =	\left[
	\begin{array}{ccc}
	 (-,-)&(-,-)&(+,+)\\
	 (-,-)&(-,-)&(+,+)\\
	 (+,+)&(+,+)&(-,-)
	\end{array}
	\right],
 \nt\\
	G_{\mu}
 &=	\left[
	\begin{array}{ccc}
	 (+,+)&(+,+)&(+,+)\\
	 (+,+)&(+,+)&(+,+)\\
	 (+,+)&(+,+)&(+,+)
	\end{array}
	\right],
 ~~
	G_y
 =  \label{BC}
	\left[
	\begin{array}{ccc}
	 (-,-)&(-,-)&(-,-)\\
	 (-,-)&(-,-)&(-,-)\\
	 (-,-)&(-,-)&(-,-)
	\end{array}
	\right],  
\end{align}
where (+,+) etc. stand for $Z_2$ parities at fixed points $y=0, \pi R$.
We can see that the gauge symmetry is explicitly broken as $SU(3) \to SU(2) \times U(1)$
by the boundary conditions. The gauge fields with $Z_2$ parities $(+,+)$ and $(-,-)$ are mode-expanded 
by use of mode functions, which are just trigonometric functions, i.e. $S_n (y) = \frac{1}{\sqrt{\pi R}} \sin (M_n y)$ 
and $C_n (y) = \frac{1}{\sqrt{\pi R}} \cos (M_n y) \ (n \neq 0), \ C_0 (y) = \frac{1}{\sqrt{2\pi R}}$, respectively.

The $Z_2$ parities of fermions are assigned for each component of the representations as follows: 
\begin{align} 
	\Psi^i({\bf 3})
 &= \big(Q_{3L}^i(+,+)+Q_{3R}^i(-,-)\big)
	\oplus \big(d_L^i(-,-)+d_R^i(+,+)\big),
	\nt\\
	\Psi^i(\bar{\bf 6})
 &= \label{BC2}
	\big(\Sigma^i_L(-,-) + \Sigma^i_R(+,+)\big)
	\oplus \big(Q_{6L}^i(+,+)+Q_{6R}^i(-,-)\big) \oplus \big(u^i_L(-,-)+u^i_R(+,+)\big)
\end{align}
where $Q_{3}^i$ and $Q_6^i$ are $SU(2)$ doublets and $d^i$ and $u^i$ are $SU(2)$ singlets. 
$\psi^i(\bar {\bf 6})$ also contain $SU(2)$ triplet exotic states $\Sigma^i$ written in a form of $2 \times 2$ symmetric matrix \cite{BN}.
In this way a chiral theory is realized in the zero mode sector by $Z_2$ orbifolding.  


The fermions are also expanded
by an ortho-normal set of mode functions. Here we will focus on the zero-mode sector, which are necessary for the argument of flavor mixing: 
\begin{align}
	\psi^i({\bf 3})
 &= \left[
	\begin{array}{c}
	 Q_{3L}^if_L^i(y) \\[6pt]
	 d_R^if_R^i(y)
	 \\
	\end{array}
	\right], \\
	\label{psi6}
	\psi^i\big(\bar {\bf 6}\big)
 &= \left[
	\begin{array}{c|c}
		\big(i\sigma^2\big)\Sigma^i\big(i\sigma^2\big)^{\!\T}
	 &  \frac1{\sqrt2}\big(i\sigma^2\big)Q_6^i\\[6pt]
	 \hline
		\frac1{\sqrt2}(Q_6^i)^{\T}\big(i\sigma^2\big)^{\!\T}
	 &  u^i
	\end{array}
	\right]
\end{align}
where $\sig$ denotes an $SU(2)$ invariant anti-symmetric tensor
$\big(\sig\big)^{\alpha\beta}\! = \epsilon^{\alpha\beta}$.
The zero mode sector of each component of $\psi^i(\bar{{\bf 6}})$ is written in terms of the same mode functions as in the case of $\psi^i({\bf 3})$.
\begin{equation}
	\Sigma^i
  = \Sigma_R^if_R^i(y)\ ,\qquad
	Q_6^i
  = Q_{6L}^if_L^i(y)\ ,\qquad
	u^i
  = u_{R}^if_R^i(y)\ .
\end{equation}
The mode function for the zero mode of each chirality is given in \cite{ALM3}: 
\begin{align}
	f^{i}_L(y)
 &= \sqrt{\frac{M^i}{1-\e ^{-2\pi RM^i}}}\e^{-M^i|y|},\quad
	f^{i}_R(y)
  = \sqrt{\frac{M^i}{\e^{2\pi RM^i}-1}}\e^{M^i|y|}. 
\end{align}

We notice that there are two left-handed quark doublets $Q_{3L}$ and $Q_{6L}$
per generation in the zero mode sector, which are massless before electro-weak symmetry breaking. 
In the one generation case, for instance, one of two independent linear combinations of these doublets should correspond
to the quark doublet in the standard model, but the other one should be regarded as an exotic state. 
Moreover, we have an exotic fermion $\Sigma_R$.  
We therefore introduce brane localized four dimensional Weyl spinors 
to form $SU(2) \times U(1)$ invariant brane localized Dirac mass terms
in order to remove these exotic massless fermions from the low-energy effective theory \cite{BN, ACP}.
\begin{equation}
\begin{aligned}
	\mathcal{L}_\text{BM}
 =& \dy \sqrt{2\pi R}\,\delta(y)\bar Q_R^i(x)
	\Big\{\eta_{ij}Q_{3L}^j(x,y)+\lambda_{ij}Q_{6L}^j(x,y)\Big\}\\*
  & +\!\dy\sqrt{2\pi R}\,
	 m_{\rm BM}\delta(y-\pi R)\tr\Big\{\bar\Sigma_R^i(x,y)\chi_{L}^i(x)\Big\}
	+({\rm h.c.})
\end{aligned}
\end{equation}
where $Q_R$ and $\chi_L$ are the brane localized Weyl fermions 
of doublet and the triplet of $SU(2)$ respectively. 
The $n \times n$ matrices $\eta_{ij}, \lambda_{ij}$ and $m_{\rm BM}$ are mass parameters. 
These brane localized mass terms are introduced at opposite fixed points 
such that $Q_R(\chi_L)$ couples to $Q_{3,6L}(\Sigma_R)$
localized on the brane at $y=0~(y=\pi R)$. 
Let us note that the matrices $\eta_{ij}, \lambda_{ij}$ can be non-diagonal,
which causes the flavor mixing \cite{AKLM} \cite{BN}.

Some comments on this model are in order. 
The predicted Weinberg angle of this model is not realistic, $\sin^2 \theta_W = 3/4$. 
Possible modification is to introduce an extra $U(1)$
or the brane localized gauge kinetic term \cite{SSS}. 
However, the wrong Weinberg angle does not affect our argument,   
since our interest is  $D^0-\bar D^0$ mixing via KK gluon exchange in the QCD sector,
whose amplitude is independent of the Weinberg angle.

Second, in our model the bulk masses of fermions are generation-dependent,
but are taken as common for both $\psi^i({\bf 3})$ and $\psi^i(\bar {\bf 6})$. 
In general, the bulk masses of each representation are mutually independent 
and there is no physical reason to take such a choice.  
It would be justified if we have some Grand Unified Theory (GUT) 
where the $\bf 3$ and $\bar {\bf 6}$ representations are embedded
into a single multiplet of the GUT gauge group. 
We do not further pursue this issue in this paper. 

\section{Flavor mixing}

In the previous section we worked in the base where fermion bulk mass terms are written in a diagonal matrix in the generation space. Then the lagrangian for fermions, which includes Yukawa couplings as the gauge interaction of $A_y$ is completely diagonalized in the generation space. Thus flavor mixing does not occur in the bulk and the brane localized mass terms for the doubled doublets $Q_{3L}$ and $Q_{6L}$ is expected to lead to the flavor mixing. 
We now confirm the expectation and discuss how the flavor mixing is realized in this model. 

First, we identify the SM quark doublet 
by diagonalizing the relevant brane localized mass term, 
\begin{align}
	\dy\sqrt{2\pi R}\,\delta(y)\bar Q_R(x)
	\big[\begin{array}{ccc}
	 \eta & \lambda
	\end{array}\big]\!\!
	\left[
	\begin{array}{c}
	 Q_{3L}(x,y)\\[1pt]
	 Q_{6L}(x,y)
	\end{array}\right]\!
 &\supset
	\sqrt{2\pi R}\,\bar Q_R(x)
	\big[\begin{array}{ccc}
	 \eta f_L(0) & \lambda f_L(0)
	\end{array}\!\big]\!\!
	\left[
	\begin{array}{c}
	 Q_{3L}(x)\\[1pt]
	 Q_{6L}(x)
	\end{array}\right]\notag\\
 &= \label{diagonal}
	\sqrt{2\pi R}\, \bar Q_R'(x)
	\big[\begin{array}{ccc}
	 m_\text{diag} & \bs0_{n\times n}
	\end{array}\!\big]\!\!
	\left[\begin{array}{c}
	 Q_{\H L}(x)\\[1pt]
	 Q_{\SM L}(x)\!\!
	\end{array}\right]
\end{align}
where 
\begin{gather}
	\left[\begin{array}{cc}
	 U_1 & U_3\\[1pt]
	 U_2 & U_4
	\end{array}\right]\!\!
	\left[\begin{array}{c}
	 Q_{\H L}(x)\\[1pt]
	 Q_{\SM L}(x)
	\end{array}\right]
  = \left[\begin{array}{c}
	 Q_{3L}(x)\\[1pt]
	 Q_{6L}(x)
	\end{array}\right]\ , \quad
	U^{\bar Q} Q_R(x) 
\label{Umatrices} 
=Q_R'(x)\ ,\\[2pt]
	U^{\bar Q}
	\big[\begin{array}{ccc}
	 \eta f_L(0) & \lambda f_L(0)
	\end{array}\!\big]\!\!
	\left[
	\begin{array}{cc}
	 U_1 & U_3\\[1pt]
	 U_2 & U_4
	\end{array}
	\right] 
  = \big[\begin{array}{ccc}
	 m_\text{diag} & \bs0_{n\times n}
	\end{array}\!\big]\ .
\end{gather}
In eq. (\ref{diagonal}),
$\eta f_L(0)$ is an abbreviation of a $n \times n$ matrix
whose $(i, j)$ element is given by $\eta_{ij}f_L^{j} (0)$, for instance.
$U_3$, $U_4$ are $n \times n$ matrices which indicate how the quark doublets of the SM are contained in each of $ Q_{3L}(x)$ and $Q_{6L}(x)$ and compose a $2n \times 2n$ unitary matrix together with $U_1$, $U_2$, which diagonalizes the brane localized mass matrix. The eigenstate $Q_\H$ becomes massive and decouples from the low energy processes, 
while $Q_\SM$ remains massless at this stage
and therefore is identified with the SM quark doublet. 
$U_3$ and $U_4$ satisfy the following unitarity condition:
\begin{equation}
\label{unitary_cond} 
  U_3^\dag U_3+U_4^\dag U_4={\bf 1}_{n\times n}.  
\end{equation} 
After this identification of the SM doublet,
Yukawa couplings are read off
from the higher dimensional gauge interaction of $A_y$, 
whose zero mode is the Higgs field $H(x)$:  
\begin{align}
  & \dy\!\left[-\frac g2\bar\psi^i({\bf 3})A_y^a\lambda^a\Gamma^y\psi^i({\bf 3})
	+g{\rm Tr}\Big\{
	\bar\psi^i({\bar{\bf 6}})A_y^a(\lambda^a)^\ast
	\Gamma^y \psi^i({\bar{\bf 6}})
	\Big\}
	\right]\notag\\
 \supset & \dy\!\left\{ 
	-g\bar Q^{i}_{3L}(x,y)H(x,y)d^{i}_R (x,y)
	-\sqrt2g\bar Q^{i}_{6L}(x,y)\sig H^{\ast}(x,y)u^{i}_R (x,y)
	+({\rm h.c.})\right\}\notag\\ 
 \supset &
	-g_4\!
	\left[
	\left\langle H^{\dagger}\right\rangle\bar d_R^{i}(x)
	I_{RL}^{i(00)}U_3^{ij}Q_{\SM L}^{j}(x)
	+\sqrt2\left\langle H^t \right\rangle \sig
	 \bar u_R^{i}(x)I_{RL}^{i(00)}U_4^{ij}Q_{\SM L}^{j}(x)
	\right]
	+({\rm h.c.})
\end{align}
where $g_4\equiv \frac g{\sqrt{2\pi R}}$ and
\begin{equation}
   I_{RL}^{i(00)}
 = \dy f_L^if_R^i
 = \frac{\pi RM^i}{\sinh(\pi RM^i)}\ ,
\end{equation}
which behaves as $2\pi RM^i\e^{-\pi RM^i}$ for $\pi RM^i \gg 1$,
thus realizing the hierarchical small quark masses without fine tuning of $M^i$.  
We thus know that the matrices of Yukawa coupling $g_4Y_u$ and $g_4Y_d$ are given as 
\begin{equation} 
	\label{Yukawa coupling} 
	g_4Y_u = \sqrt2g_4I_{RL}^{(00)} U_4, \qquad
	g_4Y_d = g_4I_{RL}^{(00)} U_3, 
\end{equation}
where the matrix $I_{RL}^{(00)}$ has elements
$\big(I_{RL}^{(00)}\big)_{ij} = \delta_{ij} I_{RL}^{i(00)}$. 
These matrices are diagonalized by bi-unitary transformations as in the SM
and Cabibbo-Kobayashi-Maskawa matrix is defined in a usual way.
\begin{equation}
	\label{cond_U3,U4}
	\left\{
	\begin{aligned}
	  \hat Y_d &=\text{diag}(\hat m_d,\cdots) = V_{dR}^\dag Y_d V_{dL}\\
	  \hat Y_u &=\text{diag}(\hat m_u,\cdots) = V_{uR}^\dag Y_u V_{uL}
	\end{aligned}
	\right.\ ,\quad
	  V_\text{CKM}\equiv V_{dL}^\dag V_{uL}\,
\end{equation}
where all the quark masses are normalized
by the $W$-boson mass as $\hat m_f =\frac{m_f}{M_W}$.
A remarkable point is
that the Yukawa couplings $g_4Y_u$ and $g_4Y_d$ are mutually related
by the unitarity condition eq.\,(\ref{unitary_cond}), 
on the contrary those are completely independent in the Standard Model. 
Thus if we set bulk masses of fermion to be universal among generations,
i.e. $M^1=M^2=M^3=\cdots=M^n$, 
then $I_{RL}^{(00)}$ is proportional to the unit matrix. 
In such a case,
$Y_u^\dag Y_u \propto U_4^{\dagger} U_4$ and $Y_d^\dag Y_d \propto U_3^{\dagger} U_3$
can be simultaneously diagonalized 
because of the unitarity condition eq.\,(\ref{unitary_cond}).
This means that the flavor mixing disappears in the limit of universal bulk masses,
as was expected in the introduction. 
In reality, off course the bulk masses should be different
to explain the variety of quark masses
and therefore the flavor mixing does not vanish.    

For an illustrative purpose to confirm the mechanism of flavor mixing,
let us consider the two generation case. 
We will see how the realistic quark masses and mixing are reproduced. 
For simplicity, we ignore CP violation and assume that $U_3, \ U_4$ are real.  
Let us notice that from the unitarity condition shown in (\ref{unitary_cond}), 
$U_3^\dag U_3+U_4^\dag U_4 = {\bf 1}_{2\times 2}$, $2 \times 2$ matrices $U_{3,4}$ can be parametrized without loss of generality as 
\begin{align} 
	\label{parametrization}
	U_4
  = \!\left[
	\begin{array}{cc}
	 \cos \theta' & -\sin \theta' \\
	 \sin \theta' & \cos \theta' \\
	\end{array}
	\right]\!\!
	\left[
	\begin{array}{cc}
	 a & 0 \\
	 0 & b \\
	\end{array}
	\right], \quad 
	U_3
  = \!\left[
	\begin{array}{cc}
	 \cos \theta & -\sin \theta \\
	 \sin \theta & \cos \theta \\
	\end{array}
	\right]\!\!
	\left[
	\begin{array}{cc}
	 \sqrt{1-a^2} & 0 \\
	 0 & \sqrt{1-b^2} \\
	\end{array}
	\right]. 
\end{align}
Actually the most general forms of $U_{3}$ and $U_{4}$ have a common orthogonal matrix
multiplied from the right, being consistent with (\ref{unitary_cond}).
The common orthogonal matrix, however, can be eliminated by suitable unitary transformation among the members of 
$Q_{\SM L}(x)$. 

Let us note that if we wish, instead of the base where bulk mass term is diagonalized, we can move to another base where 
$\theta = \theta' = 0$ by suitable unitary transformations of $Q_3$ and $Q_6$. Then in this base the bulk mass term is no longer diagonal in the generation space unless bulk masses are degenerate, and the off-diagonal elements lead to flavor mixing. In the specific case of degenerate bulk masses, the bulk mass term is still diagonal and flavor mixing disappears. 
This is another proof of why flavor mixing disappears for degenerate bulk masses.

The integral $I_{RL}^{(00)}$ is parametrized as follows. 
\begin{align} 
	I_{RL}^{(00)}
  = \left[
	\begin{array}{cc}
	 c & 0 \\
	 0 & d \\
	\end{array}
	\right]. 
	\label{parametrization2}
\end{align}
Now physical observables $\hat m_u, \hat m_c, \hat m_d, \hat m_s$ 
and the Cabibbo angle $\theta_{c}$ are all written
in terms of $a,b,c,d$ and $\theta, \theta'$. 
Namely trivial relations
\begin{alignat}{2}
	{\rm det} \big(\hat Y_d^\dagger \hat Y_d\big)
 &= \hat m_d^2 \hat m_s^{2}\ ,
 & {\rm det}\big(\hat{Y}_u^\dagger \hat{Y}_u \big)
 &= \hat m_u^2 \hat m_c^{2}\ ,\\
	{\rm Tr}\big(\hat{Y}_d^\dagger \hat{Y}_d \big)
 &= \hat{m}_d^2 + \hat{m}_s^2\ ,\qquad
 &  {\rm Tr} \big(\hat Y_u^\dagger \hat Y_u\big)
 &= \hat{m}_u^2 + \hat{m}_c^2
\end{alignat}
provide through eqs.\,(\ref{Yukawa coupling})-(\ref{parametrization2})
with
\begin{align}
	\label{cond1}
	\hat{m}_d^2 \hat{m}_s^2
 &= \big(1-a^{2}\big)\!\big(1-b^{2}\big) c^{2} d^{2},\\
	\label{cond2}
	\hat{m}_d^2 + \hat{m}_s^2
 &= \big(1-a^2\big)c^2 + \big(1-b^2\big)d^2
	+\big(a^2 - b^2\big)\!\big(c^2 - d^2\big) \sin^{2} \theta,\\
	\label{cond3}
	\hat{m}_u^2 \hat{m}_c^2
 &= 4a^{2} b^{2} c^{2} d^{2},\\
	\label{cond4} 
	\hat{m}_u^2 + \hat{m}_c^2
 &= 2\Big\{a^2c^2+b^2d^2-\big(a^2-b^2\big)\!\big(c^2-d^2\big)\sin^2\theta'\Big\}\ .
\end{align}
We also note that $\theta_c$ is given as 
\begin{align} 
	\label{Cabibbo}
 &  \tan 2\theta_c
  = \frac{\tan2\theta_{dL}-\tan2\theta_{uL}}{1+\tan2\theta_{dL}\tan2\theta_{uL}}\ ,\\
	\label{Cabibbo1}
 &  \tan 2\theta_{dL}
  = \frac{2\sqrt{(1-a^2)(1-b^2)}(d^2-c^2)\sin\theta\cos\theta}
		 {(1-a^2)(c^2\cos^2\theta+d^2\sin^2\theta)
		 -(1-b^2)(c^2\sin^2\theta+d^2\cos^2\theta)}\ ,\\   
	\label{Cabibbo2}  
 &  \tan2\theta_{uL}
  = \frac{2ab(d^2-c^2)\sin\theta'\cos\theta'}
		 {a^2(c^2\cos^2\theta'+d^2\sin^2\theta')
		 -b^2(c^2\sin^2\theta'+d^2\cos^2\theta')}
\end{align}
where angles $\theta_{dL}, \ \theta_{uL}$ are angles
parametrizing $V_{dL}, \ V_{uL}$, respectively. 
Note that five physical observables are written in terms of six parameters,
$a,b,c,d$ and $\theta, \theta'$. 
So our theory has one degree of freedom which cannot be determined by the observables.
We choose $\theta'$ as a free parameter.
Then once we choose the value of $\theta'$,
other 5 parameters can be completely fixed by the observables,
by solving eqs.\,(\ref{cond1})-(\ref{Cabibbo2})
numerically for $a,b,c,d$ and $\theta$.
The result is shown in Table 1.\footnote{
Note that $\sin\theta'$ has the upper and lower limits.
As $\sin\theta'$ goes to $\pm1$,
the bulk mass of the second generation $M^2$ goes to 0 (i.e. $d = 1$).
When $\sin\theta'$ takes a value beyond these limits, we have no solution.
}

\begin{figure}[h]
\centering
\setlength{\extrarowheight}{1pt}
\begin{tabular}{|c||c|c|c|c|c|}
 \hline
 $\sin\theta'$ & $a^2$ & $b^2$ & $c^2$ & $d^2$ & $\sin\theta$ \\ \hline\hline
 -0.9999 & 0.000015 & 0.999998 & 3.94$\tt^{-9}$ & \bf 1 & -0.00016 \\ \hline
 -0.8 & 0.0463 & 0.9951 & 4.07$\tt^{-9}$ & 3.22$\tt^{-4}$ & -0.00383 \\ \hline
 -0.6 & 0.0770 & 0.9916 & 4.19$\tt^{-9}$ & 1.88$\tt^{-4}$ & 0.00195 \\ \hline
 -0.4 & 0.0959 & 0.9894 & 4.31$\tt^{-9}$ & 1.47$\tt^{-4}$ & 0.00992 \\ \hline
 -0.2 & 0.1052 & 0.9882 & 4.43$\tt^{-9}$ & 1.31$\tt^{-4}$ & 0.01845 \\ \hline
  0.0 & 0.1062 & 0.9881 & 4.55$\tt^{-9}$ & 1.26$\tt^{-4}$ & 0.02649 \\ \hline
 0.2 & 0.0997 & 0.9889 & 4.68$\tt^{-9}$ & 1.31$\tt^{-4}$ & 0.03315 \\ \hline
 0.4 & 0.0860 & 0.9906 & 4.80$\tt^{-9}$ & 1.48$\tt^{-4}$ & 0.03745 \\ \hline
 0.6 & 0.0650 & 0.9930 & 4.93$\tt^{-9}$ & 1.89$\tt^{-4}$ & 0.03806 \\ \hline
 0.8 & 0.0365 & 0.9962 & 5.07$\tt^{-9}$ & 3.27$\tt^{-4}$ & 0.03239 \\ \hline
 0.9999 & 0.000012 & 0.999998 & 5.23$\tt^{-9}$ & \bf 1 & 0.00064\\ \hline
\end{tabular}\\[1\baselineskip]
Table 1:
Numerical result for the relevant parameters fixed by quark masses and Cabibbo angle.  
\end{figure}
 
Thus we have confirmed that observed quark masses
and flavor mixing angle can be reproduced in our model of GHU. 
Let us note that in eq.\,(\ref{Cabibbo}) 
Cabibbo angle $\theta_c$ vanishes in the limit of universal bulk mass, 
i.e. $M^1 = M^2$ leads to $c = d$ as is expected.

\section
[$D^0-\bar D^0$ mixing]
{$\bs{D^0-\bar D^0}$ mixing}

In this section,
we apply the results of the previous section to a representative FCNC process, 
$D^0-\bar D^0$ mixing responsible for the mass difference of 
two neutral $D$ mesons.\footnote{For the studies of $D^0-\bar{D}^0$ mixing 
in other new physics models, see for instance \cite{GGNP} \cite{CFW}.}

We focus on the FCNC processes of zero mode up-type quarks
due to gauge boson exchange at the tree level.
First let us consider the processes with the exchange of zero mode gauge bosons. 
If such type of diagrams exist with a sizable magnitude, 
it will easily spoil the viability of the model.

Concerning the $Z$-boson exchange,
it is in principle possible to occur the tree-level FCNC. 
Since the mode function of the zero mode gauge boson is $y$-independent, 
the overlap integral of mode functions is universal, i.e. generation independent,
just as the kinetic term of fermions are.
Thus the gauge coupling of zero mode gauge boson depends
on only the relevant quantum numbers such as the third component of weak isospin $I_3$.
Therefore the condition proposed by Glashow-Weinberg \cite{GW}
to guarantee natural flavor conservation for the theories of 4D space-time is relevant.
Note that we have two right-handed up-type quarks
belonging to $\psi(\bar{\bs6})$, $SU(2)$ singlet $u^i$
and a member of $SU(2)$ triplet $\Sigma^i$ in (\ref{psi6}).
They have different $I_3$, i.e. 0 and 1,
while they have the same electric charge and chirality.
Thus, the condition of Glashow-Weinberg is not satisfied
in the up-type quark sector and FCNC process
due to the exchange of the zero mode $Z$-boson arises at the tree level.\footnote{
The FCNC due to the exchanges of zero mode photon and gluon trivially vanish 
because the fermions of our interest have the same electric charge and color.}
However, the triplet $\Sigma^i$ is an exotic fermion
and acquires large $SU(2)$ invariant brane mass.
Thus the mixing between $u^i$ and $\Sigma^i$ is inversely suppressed
by the power of $m_{\rm BM}$ and the FCNC vertex of $Z$-boson can be safely neglected.
We may say that the condition of Glashow-Weinberg is satisfied in a good approximation 
in the processes via the zero mode gauge boson exchange.
Furthermore, the contribution by the weak gauge boson exchange is expected
to be small compared with that by the gluon exchange.

Hence, the remaining possibility is the process
via the exchange of non-zero KK gauge bosons.
In this case, the mode functions of KK gauge bosons are $y$-dependent 
and their couplings to fermions are no longer universal  
because of non-degenerate bulk masses, even if the condition of Glashow-Weinberg is met.    

Therefore, such progresses lead to FCNC at the tree level.  
In our previous paper \cite{AKLM}, 
we have calculated $K^0-\bar{K}^0$ mixing via the non-zero KK gluon exchange at the tree level 
and obtained a lower bound of the compactification scale as the prediction of our model. 

Along the same line of the argument as in our previous paper,
we here study $D^0-\bar D^0$ mixing in the up-type quark sector caused
by the non-zero KK gluon exchange at the tree level 
as the dominant contribution to this FCNC process. 

For such purpose, we derive the four dimensional effective strong interaction vertices 
with respect to the zero modes of up-type quarks relevant for our calculation: 
\begin{align}
	\mathcal{L}_\s
 \supset\,&
	\frac{g_\s}{2\sqrt{2\pi R}}G_\mu^{a}
	\Big(
	\bar u^i_R\gm\lambda^a u^i_R
	+\bar Q^{i}_{3L}\lambda^a\gm Q^{i}_{3L}
	+\bar Q^{i}_{6L}\lambda^a\gm Q^{i}_{6L}
	\Big)\notag\\*
 &  +\frac{g_\s}2 G_{\mu}^{a(n)}\!
	\left\{
	\bar u^i_R\lambda^a\gm u^i_R I_{RR}^{i(0n0)}
	+(-1)^n
	\big(
	\bar Q^{i}_{3L}\lambda^a\gm Q^{i}_{3L}
	+\bar Q^{i}_{6L}\lambda^a\gm Q^{i}_{6L}
	\big)
	I_{RR}^{i(0n0)}
	\right\}\notag\\
 \supset\,&
	\frac{g_\s}{2\sqrt{2\pi R}}G_\mu^{a}\!
	\left(
	\bar{\tilde u}_R^i\gm\lambda^a\tilde u_R^i
	+\bar {\tilde u}_L^i\lambda^a\gm\tilde u_L^i
	\right)\notag\\*
 &  +\frac{g_\s}{2} G_\mu^{a(n)}
	\bar{\tilde u}_R^i\lambda^a\gm\tilde u_R^j\!
	\left(V_{uR}^\dag I_{RR}^{(0n0)}V_{uR}\right)_{ij}\notag\\*
 &  \label{strong}
	+\frac{g_\s}2 G_\mu^{a(n)}
	\bar{\tilde u}^i_L\lambda^a\gm\tilde u_L^j
	(-1)^n\!
	\left(
	V_{uL}^\dag U_3^\dag I_{RR}^{(0n0)}U_3V_{uL}
	+V_{uL}^\dag U_4^\dag I_{RR}^{(0n0)}U_4V_{uL}
	\right)_{ij}
\end{align}
where $I_{RR}^{i(0n0)}$ is a overlap integral relevant for gauge interaction,     
\begin{equation} 
   \label{vertexfunction} 
   I_{RR}^{i(0n0)}
 = \frac1{\sqrt{\pi R}}\dy\big(f^i_R\big)^2\cos(M_n y)
 = \frac1{\sqrt{\pi R}}
   \frac{4(M^i)^2}{4(M^i)^2+\left(\frac nR\right)^2}
   \frac{(-1)^n\e^{2\pi M^i R}-1}{\e^{2\pi M^i R}-1}.    
\end{equation}

Let us note
that the overlap integrals for left-handed fermion $I_{LL}^{i(0n0)}$ 
are related to those for the right-handed ones $I_{RR}^{i(0n0)}$ 
as $I_{LL}^{i(0n0)} = (-1)^{n} I_{RR}^{i(0n0)}$ 
since the chirality exchange corresponds to the exchange of two fixed points. 
In eq.\,(\ref{strong}), 
$\tilde u$ denotes mass eigenstates,
$\big(\tilde u^1, \tilde u^2\big) = (u, c)$.
We can see from (\ref{strong})
that the FCNC appears in the couplings of non-zero KK gluons
due to the fact that $I_{RR}^{(0n0)}$ is not proportional
to the unit matrix in the generation space, while the coupling of the zero mode gluon is flavor conserving, as we expected.

\begin{figure}[h]
\[
\begin{array}{ccc}
\includegraphics[bb= 0 0 105 66, scale=1]{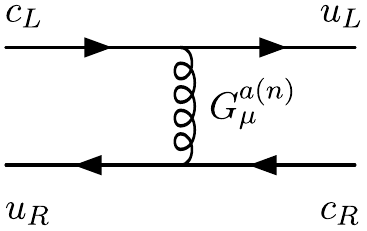} \qquad&\qquad
\includegraphics[bb= 0 0 104 66, scale=1]{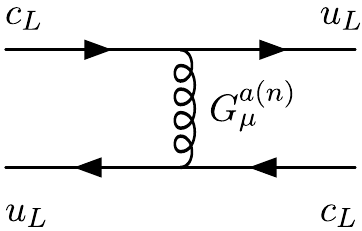} \qquad&\qquad
\includegraphics[bb= 0 0 105 65, scale=1]{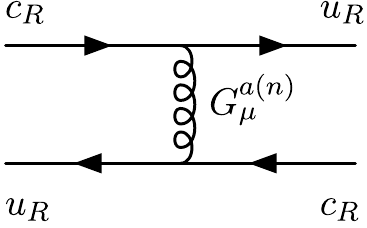}\\[8pt]
\text{(i) LR type} \qquad&\qquad
\text{(ii) LL type} \qquad&\qquad
\text{(iii) RR type}
\end{array}
\]
\caption{The diagrams of $D^0-\bar D^0$ mixing via KK gluon exchange}
\label{fig1}
\end{figure}

The Feynman rules necessary for the calculation of $D^0-\bar D^0$ mixing
can be read off from (\ref{strong}).
\begin{align}
	\begin{array}{c}
	\includegraphics[bb= 0 0 108 38]{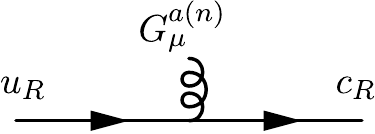}
	\end{array}
 &= \frac{g_\s}{2}\!
	\left(V_{uR}^\dag I_{RR}^{(0n0)}V_{uR}\right)_{21}\!
	\lambda^a\gm R,\\[5pt]
    \label{Lvertex}
	\begin{array}{c}
	\includegraphics[bb= 0 0 107 39]{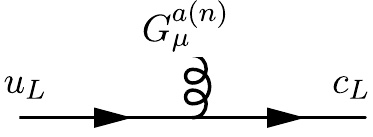}
	\end{array}
 &= \frac{g_\s}{2} (-1)^n\!
	\left(
	V_{uL}^\dag U_3^\dag I_{RR}^{(0n0)}U_3 V_{uL}
	+V_{uL}^\dag U_4^\dag I_{RR}^{(0n0)}U_4V_{uL}
	\right)_{21}\!\!
	\lambda^a \gm L,\\
	\label{propagator}
	\begin{array}{c}
	\includegraphics[bb= 0 0 119 31]{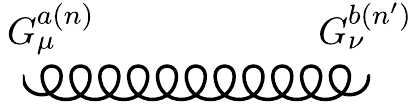}\!\!\!\!
	\end{array}
 &= \delta_{nn'}\delta_{ab}\frac{\eta_{\mu\nu}}{k^2-M_n^2}
	\qquad{\big(\,\text{'t Hooft-Feynman gauge}\,\big).}
\end{align}
The non-zero KK gluon exchange diagrams,
which give the dominant contribution to the process of $D^0-\bar D^0$ mixing,
are depicted in Fig.\,\ref{fig1}.

Note that in the case of the $K^0-\bar K^0$ mixing
which is given by similar diagrams to those in Fig.\,\ref{fig1} (see appendix),
only the `LR type' diagram was considered in the calculation,
by the reason that the hadronic matrix element
of LR type effective 4-Fermi lagrangian is relatively enhanced
with a factor $\frac{m_K}{m_d+m_s}$ \cite{AKLM}
compared to the matrix elements of LL and the RR type effective lagrangian.  
In the case of the $D^0-\bar D^0$ mixing,
the factor $\frac{m_D}{m_u+m_c}$ is not so large
and in addition to the LR type diagram we calculate LL and RR type diagrams as well.\footnote{
It turns out that even in the case of $K^0-\bar K^0$ mixing,
the LL and RR type processes is not less important
and even give dominant contribution, as we will see in appendix.
In our previous paper \cite{AKLM}, we naively assumed in the analysis
that the mode sums for the LL and the RR type diagrams are comparable to
that for the LR type diagram, but it turns out not to be the case.} 

By noting the fact $k^2\ll M_n^2$ for $n \neq 0$,
with $M_n = n/R$ being the mass of $n$-th KK gluon and $k^{\mu}$ being external momentum,
the contribution of LR diagram of Fig.\,\ref{fig1} is written
in the form of effective four-Fermi lagrangian
obtained by use of Feynman rules listed above,  
\begin{align}
	\begin{array}{c}
	\includegraphics[bb= 0 0 105 66]{figure/KKgluonexchangeLR.pdf}
	\end{array}
 \sim &
	\sum_{n=1}^\infty
	\frac{g_\s^2}4\frac{(-1)^n}{M_n^2}\!
	\left(
	V_{uL}^\dag U_3^\dag I_{RR}^{(0n0)}U_3V_{uL}
	+V_{uL}^\dag U_4^\dag I_{RR}^{(0n0)}U_4 V_{uL}
	\right)_{21}\notag\\[-0.7\baselineskip]
 &  \label{DDbarLR}
	\times
	\left(
	V_{uR}^\dag I_{RR}^{(0n0)}V_{uR}
	\right)_{21}\!
	\big(\bar u_L\lambda^a \gamma^\mu c_L\big)\!
	\big(\bar u_R\lambda^a \gamma_\mu c_R\big).
\end{align}
Similarly, the LL and the RR type diagrams of Fig.\,\ref{fig1} give
\begin{align}
	\begin{array}{c}
	\includegraphics[bb= 0 0 104 66]{figure/KKgluonexchangeLL.pdf}
	\end{array}
 \sim &
	-\sum_{n=1}^\infty
	\frac{g_\s^2}4\frac1{M_n^2}\!
	\left(
	V_{uL}^\dag U_3^\dag I_{RR}^{(0n0)}U_3V_{uL}
	+V_{uL}^\dag U_4^\dag I_{RR}^{(0n0)}U_4 V_{uL}
	\right)_{21}^2\notag\\[-0.6\bls]
 &  \label{DDbarLL}
	\times
	\big(\bar u_L\lambda^a \gamma^\mu c_L\big)\!
	\big(\bar u_L\lambda^a \gamma_\mu c_L\big),\\[0.5\bls]
	\begin{array}{c}
	\includegraphics[bb= 0 0 105 65	]{figure/KKgluonexchangeRR.pdf}
	\end{array}
 \sim & \label{DDbarRR}
	-\sum_{n=1}^\infty
	\frac{g_\s^2}4\frac1{M_n^2}\!
	\left(
	V_{uR}^\dag I_{RR}^{(0n0)}V_{uR}
	\right)_{21}^2\!
	\big(\bar u_R\lambda^a \gamma^\mu c_R\big)\!
	\big(\bar u_R\lambda^a \gamma_\mu c_R\big).
\end{align} 
The sum over the integer $n$ is convergent
and the coefficients of the effective lagrangian
(\ref{DDbarLR})-(\ref{DDbarRR}) are suppressed by the compactification scale 
as $1/M_{\rm c}^2$ where $M_{\rm c} = R^{-1}$. 
We can verify, as we expect, that the coefficient vanishes in the limit of universal bulk masses 
$M^1 = M^2 = \cdots$ by use of the unitarity condition (\ref{unitary_cond}), since $I_{RR}^{(0n0)}$ is proportional to the unit matrix in this limit;
\begin{align}
	V_{uL}^\dag\!
	\left(
	U_3^\dag I_{RR}^{(0m0)}U_3
	+U_4^\dag I_{RR}^{(0m0)}U_4
	\right)\! V_{uL}
 &~\xrightarrow{M^1 = M^2 =\, \cdots\,}~
	V_{uL}^\dag\!
	\left(U_3^\dag U_3+U_4^\dag U_4\right)\!
	V_{uL}I_{RR}^{(0m0)}
 \propto {\mathbf 1}_{n\times n}\ ,\notag\\
	V_{uR}^\dag I_{RR}^{(0m0)}V_{uR}
 &~\xrightarrow{M^1 = M^2 =\, \cdots\,}~
	V_{uR}V_{uR}^\dag I_{RR}^{(0m0)}
  \propto {\mathbf 1}_{n\times n}\ .
\end{align}

Comparing the calculation of (\ref{DDbarLR})-(\ref{DDbarRR}) with the experimental data,
we can obtain a lower bound on the compactification scale.
The most general effective Hamiltonian for $\Delta C=2$ processes
due to some ``new physics" at a high scale $\Lambda_\text{NP} \gg M_W$
can be written as follows;
\begin{equation} 
\label{effectiveH}
	\mathcal H_\text{eff}^{\Delta C=2}
  = \frac1{\Lambda_\text{NP}^2}\!
	\left(
	\sum_{i=1}^5 z_iQ_i
	+\sum_{i=1}^3 \tilde z_i\tilde Q_i
	\right)
\end{equation}
where
\begin{gather}
	Q_1
  = \bar u_L^\alpha \gamma_\mu c_L^\alpha
	\bar u_L^\beta \gamma^\mu c^\beta_L\ ,\qquad
	Q_2
  = \bar u_R^\alpha c_L^\alpha\bar u_R^\beta c^\beta_L\ ,\qquad
	Q_3
  = \bar u_R^\alpha c_L^\beta\bar u_R^\beta c^\alpha_L\ ,\notag\\*
	\label{Qi}
	Q_4
  = \bar u_R^\alpha c_L^\alpha\bar u_L^\beta c^\beta_R\ ,\qquad
	Q_5
  = \bar u_R^\alpha c_L^\beta\bar u_L^\beta c^\alpha_R\ ,
\end{gather}
and indices $\alpha, \beta$ stand for the color degrees of freedom.
The operators $\tilde Q_{1,2,3}$ are obtained from the $Q_{1,2,3}$
by the chirality exchange $L \leftrightarrow R$.
Since the contribution of the SM to the mixing is poorly known, we can get the constraint on the new physics directly from the experimental data assuming that there is no accidental cancellation between the contributions of the SM and new physics. If we assume one of these possible operators gives dominant contribution to the mixing, each coefficient is independently constrained as follows, with the constraints for $\tilde{z}_i$ are the same with those for $z_{i} \ (i = 1,2,3)$ \cite{GGNP};
\begin{gather} 
 |z_1|\leq 5.7\times 10^{-7}
 \left(\frac{\Lambda_{\rm NP}}{1{\rm TeV}}\right)^2\ ,
 \qquad
 |z_2|\leq 1.6\times 10^{-7}
 \left(\frac{\Lambda_{\rm NP}}{1{\rm TeV}}\right)^2\ ,\notag\\
 \label{constraints}
 |z_3|\leq 5.8\times 10^{-7}
 \left(\frac{\Lambda_{\rm NP}}{1{\rm TeV}}\right)^2\ ,
 \qquad
 |z_4|\leq 5.6\times 10^{-8}
 \left(\frac{\Lambda_{\rm NP}}{1{\rm TeV}}\right)^2\ ,\\
 |z_5|\leq 1.6\times 10^{-7}
 \left(\frac{\Lambda_{\rm NP}}{1{\rm TeV}}\right)^2\ \notag
\end{gather}
where $\Lambda_\text{NP}$ is regarded as the compactification scale in our case. 
All we have to do is to represent (\ref{DDbarLR})-(\ref{DDbarRR})
by use of (\ref{Qi}) and to utilize these constraints (\ref{constraints}).

We can rewrite the LR type effective lagrangian (\ref{DDbarLR}) in terms of scalar type effective Hamiltonian by using the Fiertz transformation;
\begin{equation}
	\label{DDbarLR2}
	-\frac{\pi\alpha_\s}2
	(\alpha_u+\alpha'_u)
	\sin2\theta_{uR} R^2
	S_\text{KK}^{LR}
	\left(4Q_4-\frac43Q_5\right)
\end{equation}
where we use a following relation about Gell-Mann matrices $\lambda^a$ 
normalized as $\tr\,(\lambda^a\lambda^b) = 2\delta^{ab}$;
\begin{equation}
	\sum^8_{a=1}
	(\lambda^a)_{\alpha\beta}
	(\lambda^a)_{\gamma\delta}
  = 2\delta_{\alpha\delta}\delta_{\beta\gamma}
	-\frac23\delta_{\alpha\beta}\delta_{\gamma\delta}
\end{equation}
and the four-dimensional $\alpha_\s$ is defined by
\begin{equation}
	\alpha_\s
  = \frac{(g_\s^{4D})^2}{4\pi}
  = \frac1{2\pi R}\frac{g_\s^2}{4\pi}.
\end{equation}
Parameters in (\ref{DDbarLR2}) are defined as follows:
\begin{align}
	\alpha_u
 \equiv& \label{alpha_u}
	-\!(1-a^2)\sin2\theta_{uL} \cos^2\!\theta
	+(1-b^2)\sin2\theta_{uL} \sin^2\!\theta\notag\\*
 &  -\!\sqrt{(1-a^2)(1-b^2)}\cos2\theta_{uL} \sin2\theta\ ,\\[2pt]
	\alpha'_u
 \equiv& \label{alphauprime}
	-a^2\sin2\theta_{uL} \cos^2\!\theta'+b^2\sin2\theta_{uL} \sin^2\!\theta'
	-ab\cos2\theta_{uL} \sin2\theta'\ ,\\[2pt]
	S_\text{KK}^{LR}
 \equiv&~
	\pi R\sum^\infty_{n=1}
	\frac{(-1)^n}{n^2}\Big(I_{RR}^{1(0n0)}-I_{RR}^{2(0n0)}\Big)^2
\end{align}
and $\theta_{uR}$ is an angle in the rotation matrix $V_{uR}$
to diagonalize $I_{RL}^{(00)}U_4 U_4^\dag I_{RL}^{(00)}$:
\begin{equation}
   \tan2\theta_{uR} 
 = \frac{2(a^2-b^2) cd \sin\theta'\cos\theta'}
   {c^2(a^2\cos^2\theta' + b^2\sin^2\theta')
    -d^2(a^2\sin^2\theta' + b^2\cos^2\theta')}.
\end{equation}
The constant $\alpha_{\s}$ should be estimated at the scale $\mu_D = 2.8$\,GeV
where the $\Delta C = 2$ process is actually measured \cite{UTfit}.
So we have to take into account the renormalization group effect from the weak scale 
down to $\mu_D$:
\begin{equation}
	\label{alphasmuD}
	\alpha_\s^{-1}(\mu_D)
  = \alpha_\s^{-1}(M_Z)
	-\frac1{6\pi}\!
	\left(
	23\ln\frac{M_Z}{m_b} + 25\ln\frac{m_b}{\mu_D}
	\right)
  \quad \lra \quad
  \alpha_\s(\mu_D) \approx 0.240
\end{equation}
where $\alpha_\s(M_Z) \approx 0.1184$ has been put. 

Similarly, the effective Hamiltonian of LL type (\ref{DDbarLL}) and the RR type (\ref{DDbarRR})
are respectively rewritten;
\begin{equation} 
\label{nonchiralflip}
	-\frac{\pi\alpha_\s}2
	(\alpha_u+\alpha'_u)^2 R^2
	S_\text{KK}^{LL}
	\cdot\frac43 Q_1\ ,\qquad
	-\frac{\pi\alpha_\s}2
	\sin^2\!2\theta_{uR} R^2
	S_\text{KK}^{RR}
	\cdot\frac43 \tilde Q_1\ ,
\end{equation}
where
\begin{equation}
   S_\text{KK}^{LL}
 = S_\text{KK}^{RR}
 = \pi R\sum^\infty_{n=1}
   \frac1{n^2}\Big(I_{RR}^{1(0n0)}-I_{RR}^{2(0n0)}\Big)^2\ .
\end{equation}

\begin{figure}[t]
\begin{equation}
\begin{array}{cc}
   \includegraphics[bb=0 0 600 480, scale=0.43]{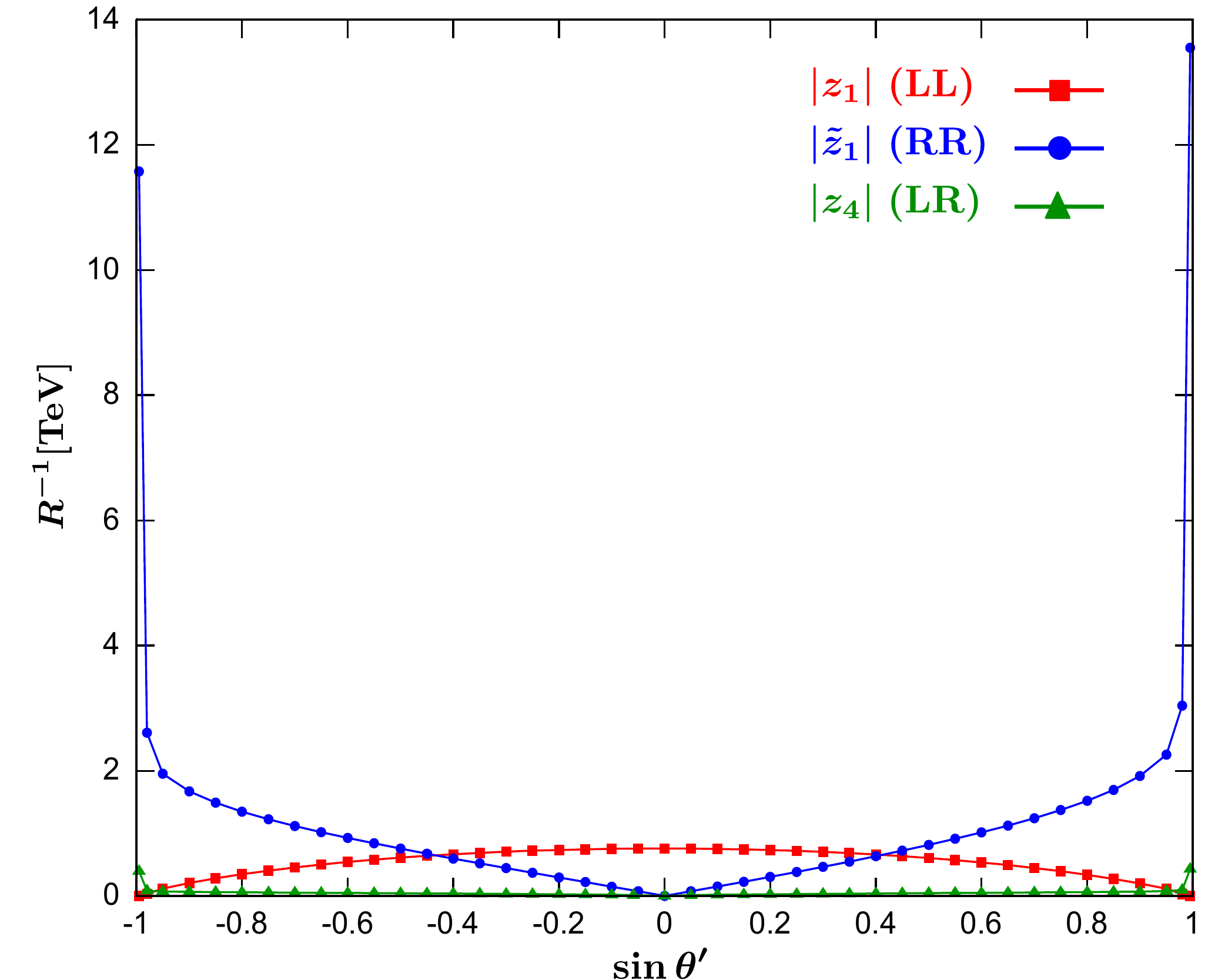}
\end{array}\notag
\end{equation}
\caption{Lower bounds on $R^{-1}$ as a function of $\sin\theta'$
obtained for each of 3 types of diagrams.}
\label{fig2}
\end{figure}

Combining these results we obtain the lower bounds for the compactification scale
from the constraint (\ref{constraints}).
First let us assume that only one of the three types of diagrams (LL,  RR,  LR) gives
dominant contribution to the mixing. 
Then we get lower bound on the compactification scale by use of the upper bound
on the relevant coefficients $z_1, z'_1$ and $z_4$ given in (\ref{constraints}): 
\begin{align}
\left\{\!
\begin{array}{l}
	\dis|z_1|:~R^{-1}
  \gtrsim
	9.39\left|\alpha_u+\alpha'_u\right|\sqrt{S_\text{KK}^{LL}}
	\times10^2\,\big[\text{TeV}\big]\ ,\\[10pt]
	\dis|\tilde z_1|:~R^{-1}
  \gtrsim
	9.39\left|\sin2\theta_{uR}\right|\sqrt{S_\text{KK}^{RR}}
	\times10^2\,\big[\text{TeV}\big]\ ,\\[10pt]
	\dis|z_4|:~R^{-1}
  \gtrsim
	5.19\sqrt{\big|(\alpha_u+\alpha'_u)\sin2\theta_{uR}S_\text{KK}^{LR}\big|}
	\times10^3\,\big[\text{TeV}\big]\ .
\end{array}
\right.
\end{align} 
Let us note that LR type diagram yields both of $Q_4$ and $Q_5$ operators
as is seen in (\ref{DDbarLR2}). 
We, however, can safely ignore the contribution of $Q_5$ to the mixing,
because in (\ref{DDbarLR2}) the coefficient of the operator is smaller than that of $Q_4$
and also because the magnitude of the hadronic matrix element of $Q_4$ is known
to be greater than that of $Q_5$,
as the constraint for $z_4$ is more severe that that for $z_5$ in (\ref{constraints}).
This is why we used the constraint for $z_4$ alone to get the lower bound
for the case of LR type diagram.
Since our theory has one free parameter, say $\theta'$,
each lower bound on $R^{-1}$ depends on it.
The obtained numerical results are given in Fig.\,\ref{fig2}, 
where the lower bound on $R^{-1}$ is plotted as a function of $\sin\theta'$
for each type of diagram. 

\begin{figure}[t]
\begin{equation}
\begin{array}{cc}
   \includegraphics[bb=0 0 600 480, scale=0.43]{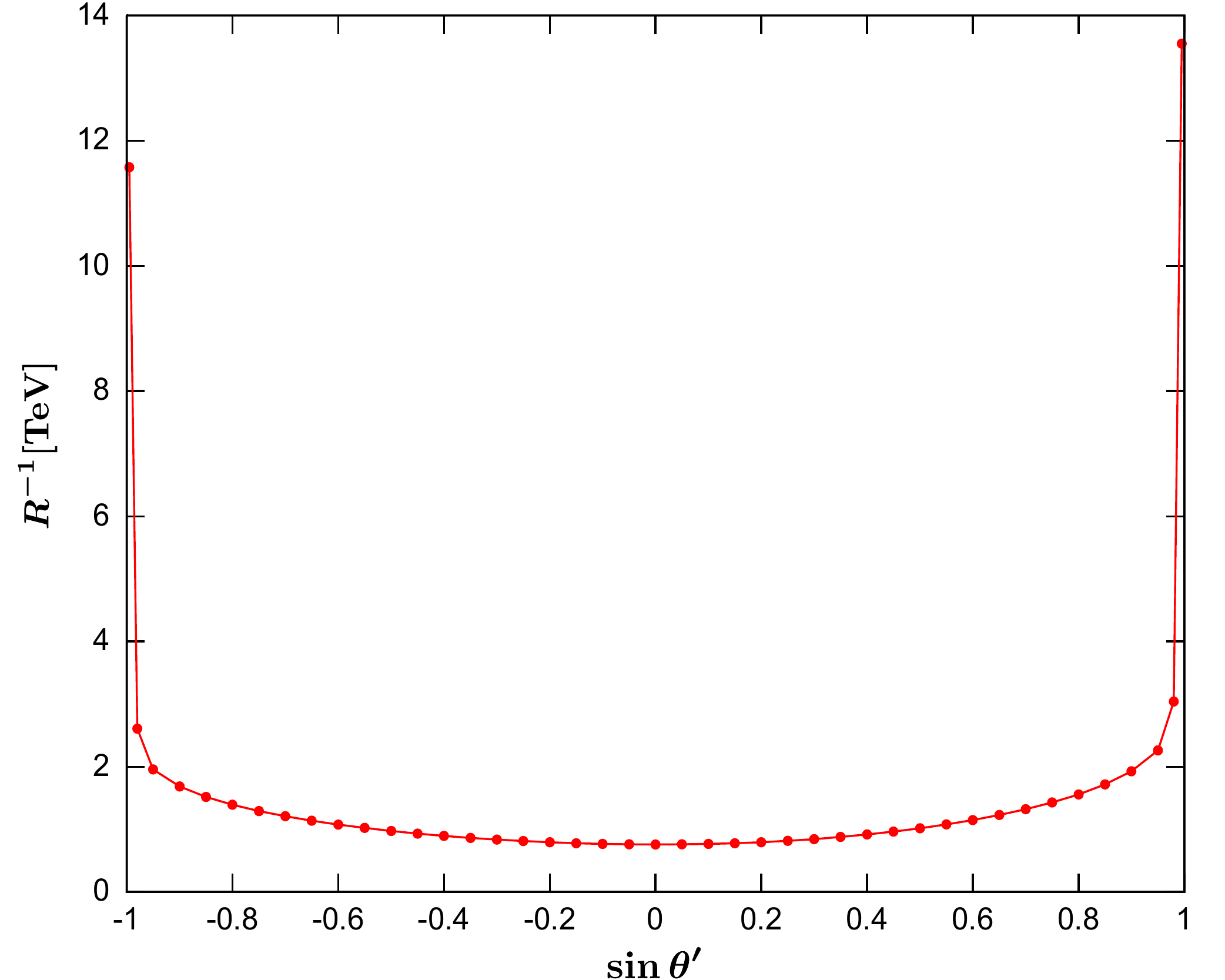}
\end{array}\notag
\end{equation}
\caption{Lower bound on $R^{-1}$}
\label{fig3}
\end{figure}

Now we learn from Fig.\,\ref{fig2}
that the contribution of LR type diagram is always negligible.
Thus we can combine the contributions to the mixing from LL and RR type diagrams together
in order to get real lower bound on the compactification scale.
Namely, by use of the fact $\big\langle\bar D^0\big|Q_1\big|D^0\big\rangle 
= \langle\bar D^0\big|\tilde Q_1\big|D^0\big\rangle$
(due to the parity symmetry of strong interaction)
we can sum up the coefficients
for $Q_1$ and $\tilde Q_1$ shown in (\ref{nonchiralflip}) together
and can utilize the constraint for $z_1$ (or equally for $\tilde z_1$)
to get the bound for $R^{-1}$: 
\begin{equation}
	R^{-1}
 \gtrsim
	9.39\sqrt{
	\Big\{(\alpha_u+\alpha'_u)^2+\sin^2 2\theta_{uR}\Big\}S_\text{KK}^{LL}}
\times10^2\,\big[\text{TeV}\big],   
\end{equation}
where $S_{\rm KK}^{LL} = S_{\rm KK}^{RR}$ is understood.
Such obtained lower bound for $R^{-1}$ is displayed in Fig.\,\ref{fig3}.

If we require that the prediction of our model is consistent with data, 
irrespectively of the choice of $\theta'$, 
we get the most stringent bound on the compactification scale,
which should be the possible largest value in Fig.\,\ref{fig3},
i.e. $R^{-1} \gtrsim 14\,{\rm TeV}$.
However, the most stringent bound comes from the very extreme case $|\sin\theta'| \simeq 1$.
For $|\sin\theta'| \lesssim 0.4$, the LL type contribution is dominant 
and provides a mild lower bound $0.8\,{\rm TeV}$. 
As for the other range of $\theta'$ except for the case $|\sin \theta'| \simeq 1$, 
the RR type contribution is dominant and the lower bound becomes around ${\cal O}$($1$\,TeV)
depending on the value of $\theta'$.

A few comments are in order.
Fig.\,\ref{fig2} indicates that the lower bounds obtained
from the LR and the RR type contributions vanish at $\sin \theta' = 0$.
In this case, we see the Yukawa coupling for up-type quarks becomes diagonal:
$\theta_{uL} = \theta_{uR} = 0$.
Thus in this extreme case the up type quark mixings disappear
and the contributions of KK gluon exchange accidentally vanish for these two types of diagram.
Note, however, that the lower bound obtained from the LL type contribution does not vanish
even though $\theta_{uL} = 0$. 
This is because 
$\theta$ in (\ref{parametrization}) relevant for down-type quark mixing 
also contributes to the left-handed FCNC current as is seen in (\ref{alpha_u}).
Namely, because of the mixing between $Q_{3L}$ and $Q_{6L}$,
$U_3$ also contributes to the FCNC vertex (\ref{Lvertex}),
which is non-diagonal even for $\theta' = 0$.
Thus even in the case of $\sin \theta' = 0$
we get a meaningful lower bound on $M_{\rm c}$ (see Fig.\,\ref{fig3}). 


\section{Suppression mechanism of FCNC} 
Note that the obtained lower bounds are smaller
than what we naively expect assuming that the tree level diagram
relevant for the FCNC process is simply suppressed by $1/M_{\rm c}^2$ \cite{UTfit}:   
\begin{align}
 \frac1{M_{\rm c}^2} \lesssim 10^{-6}\ \big[{\rm TeV}^{-2}\big]
 \quad \longrightarrow \quad
 M_{\rm c} \gtrsim {\cal O}(10^3) \,[{\rm TeV}]\ ,
\end{align}
which is much more stringent than the lower bound
we obtained except for the extreme case of $|\sin \theta'| \simeq 1$.

This apparent discrepancy may be attributed to a suppression mechanism in our scenario, 
which we will see now. The rate of $D^0-\bar D^0$ mixing is handled
by the factor $\big(I_{RR}^{1(0n0)} - I_{RR}^{2(0n0)}\big)^2$
as is seen in $S_\text{KK}^{LR}$ and $S_\text{KK}^{LL}$,
which is sensitive to the non-universality of gauge coupling of KK gluons
coming from the difference of bulk masses \big($M^1 \neq M^2 $\big).
It is easy to see that this factor is automatically suppressed
for generations with light quarks such as 1st and 2nd generations. 

Recall that in GHU hierarchical small quark masses are naturally realized
without fine tuning by exponential suppression factors $\e^{-\pi RM^{i}}$ 
with $\pi RM^{i} \gg 1$. 
On the other hand, when $\pi RM^i \gg 1$,
the ``width" $1/M^i$ of the mode function of zero-mode fermion
is much smaller than the period $\frac{2\pi R}n$
of KK gauge boson mode functions $\cos\!\big(\frac nRy\big)$.\footnote{
Note that KK modes with relatively small $n$ play an important role
in the convergent mode sum $\sum_n$.} 
Then the exponential dumping of the fermion mode functions is so fast
that the mode functions of KK gauge bosons behave as almost constant in the overlap 
integral. Thus the situation mimics the interaction vertex for zero-mode gauge boson and 
the gauge coupling of KK gauge bosons becomes almost universal. 
Therefore FCNC processes at the tree level should be automatically suppressed 
for the processes with respect to the light quarks by this mechanism.

In fact, 
analytic calculations of the mode sum in $S_\text{KK}^{LL(RR),\,LR}$ give approximate expressions for the limit $\pi R M^i \gg 1$, 
\begin{align}
	S_{\rm KK}^{LL}
 &\simeq
	\frac{\pi^2}4\frac{(\mu-\nu)^2}{\mu\nu(\mu+\nu)}\ ,\\ 
	S_{\rm KK}^{LR}
 &\simeq
	-\pi^2
	\left\{
	\frac{\e^{-\mu}+\e^{-\nu}}2
	-\frac{\mu^2+\nu^2-\mu\nu}{\mu\nu(\mu-\nu)}\!\left(\e^{-\nu}-\e^{-\mu}\right)
	\right\}\ .
\end{align}
where $\mu \equiv 2\pi RM^1$ and $\nu \equiv 2\pi RM^2$.
A remarkable thing here is that $S_\text{KK}^{LL}$ \big($S_\text{KK}^{RR} $\big) is suppressed
by an inverse power of $M^i$, while $S_\text{KK}^{LR}$ is exponentially suppressed.
Thus the suppression of FCNC is much more severe for the contribution of the LR type diagram. This is the reason why the contribution of the LR type diagram is always much smaller than those of LL and RR type diagrams for any value of $\sin\theta'$. 
We may understand qualitatively why $S_\text{KK}^{LR}$ is much smaller than $S_\text{KK}^{LL}$ as 
follows. Let us note that the mode sum in $S_\text{KK}^{LR}$,
\begin{equation}
   \sum_{n=1}^\infty\frac{(-1)^n}{n^2}
 = \sum_{n=1}^\infty\left\{\frac1{(2n)^2}-\frac1{(2n-1)^2}\right\}
 = \sum_{n=1}^{\infty} \frac{-4n+1}{(2n)^2(2n-1)^2}\ ,
\end{equation}
behaves as $\sim -\frac14 \sum\frac{1}{n^{3}}$ for larger $n$. 
Thus the convergence of $S_\text{KK}^{LR}$ is faster than that of $S_\text{KK}^{LL}$. That means in the mode sum in $S_\text{KK}^{LR}$ only smaller integer of $n$ contributes to the sum and therefore the validity of the approximation $1/M^i \ll \frac{2\pi R}n$ to ensure the universality of gauge couplings is much better in $S_\text{KK}^{LR}$ than in $S_\text{KK}^{LL}$.   

We may say that the suppression mechanism for the case of LR type contribution 
is similar to the famous GIM-mechanism
where FCNC is suppressed by a typical factor $\frac{m_s^2 - m_d^2}{M_W^2}$,
since $\e^{-\mu} \sim \frac{m_{d}^{2}}{M_{W}^{2}} \ (\mu = 2\pi RM^{1})$ etc. 
Thus it is reasonable to call this suppression mechanism ``GIM-like".\footnote{A GIM mechanism for suppression of flavor violation was known 
and well studied in warped space models \cite{RSFCNC}.}   

In the exceptional extreme case $|\sin \theta'| \simeq 1$,
the bulk mass $M^2$ happens to be relatively small
and these suppression mechanisms do not work.
That is why we get the severe lower bound on the compactification scale in the extreme case. 

\section{Summary}
In this paper,
we have discussed the flavor mixing
and the resulting flavor changing neutral current (FCNC) processes in the framework
of five dimensional $SU(3) \otimes SU(3)_{\rm color}$ gauge-Higgs unification scenario. 
As the FCNC process,
in this paper we have discussed $D^0-\bar D^0$ mixing
in the light of the recent progress in the measurements of $D^0-\bar D^0$ mixing,
which is known to be induced by the exchange of non-zero KK gluons at the tree level.

The gauge-Higgs sector in the scenario is governed by gauge principle
and therefore is very predictive, as is seen in the case of Higgs mass,
whose radiative correction is guaranteed to be finite
by the higher dimensional gauge symmetry
regardless of the non-renormalizability of the theory \cite{HIL}.
On the other hand,
understanding the flavor physics in the fermion sector is a challenging issue
since Yukawa coupling is originated from the (flavor universal) gauge coupling.

In our previous paper discussing $K^0-\bar K^0$ mixing \cite{AKLM},
we have shown that how flavor mixings are realized in the scenario.
In the present paper we have stressed that flavor mixing is due to the interplay
between the bulk masses and the brane localized masses.
Namely by suitable unitary transformations of fermion we can always take the basis
where the origin of flavor mixing is solely attributed
to either of bulk mass term or brane localized mass term.
Thus physical flavor mixing stems
only when we cannot ``diagonalize" both of bulk and brane localized mass term simultaneously.
This suggests that once bulk masses get degenerated simultaneous diagonalization becomes possible and the flavor mixing of the theory completely disappears in this limit.
We have confirmed this important property by explicit calculation.
In this way, in our model of gauge-Higgs unification the ``new physics" contributions,
namely all contributions of non-zero KK modes to FCNC,
disappear for the limit of degenerate bulk masses. 

The above argument implies that FCNC is under control
by the mass-squared difference of quarks,
just as in the case of the standard model
where $D^0-\bar D^0$ mixing is handled by $\frac{m_s^2-m_d^2}{M_W^2}$
coming from the box diagram for the simplified 2 generation case: the famous GIM mechanism.
To be more precise, the situation is a little different in these two kinds of theory.
Namely, in our model Cabibbo angle and therefore all FCNC disappear
for degenerate bulk masses, $c = d$ in (\ref{parametrization2}), even if $m_d \neq m_s$.
On the other hand, the limit $c = d$ corresponds to the limit
where averaged squared-masses of quarks get degenerated
between the first and the second generations, i.e. $m_d^2+m_u^2 = m_s^2+m_c^2$
\big(see for instance e.g. (\ref{cond1}) to (\ref{cond4})\big).
Thus FCNC due to the exchange of non-zero KK gluons at the tree level is controlled by
\begin{equation} 
	\label{massdifference}
	\frac{(m_s^2+m_c^2)-(m_d^2+m_u^2)}{M_W^2}\ ,        
\end{equation} 
while the box diagram due to the exchange of zero-mode charged gauge boson is also handled
by $\frac{m_s^2-m_d^2}{M_W^2}$ just as in the standard model.

The fact that $D^0-\bar D^0$ mixing is handled
by the mass-squared difference of quarks in our model
suggests that we have a suppression mechanism of FCNC, similar to GIM mechanism,
even for the contribution of KK gluon exchange.
In fact we have shown that the amplitude of KK gluon exchange is suppressed
concerning the contribution of light quarks of first and second generations.
The reason of the suppression was argued to be the approximate universality
of gauge couplings of KK gluons for light quarks
and the suppression also has been found to really happen in the $D^0-\bar D^0$ mixing  by explicit analytic calculations.
Interestingly, the extent of the suppression turns out
to be different for two kind of Feynman diagrams contributing to the gluon exchange,
i.e. LR type diagram ``with chirality flip"
and LL or RR type diagrams ``without chilality flip".
To be more concrete, the suppression in the LR type diagram is exponential suppression
by the factors $\e^{-2 \pi R M^1}$, $\e^{-2\pi R M^2}$,
while the suppression in the LL or RR type diagrams is only
by the inverse powers of large bulk masses, $\frac1{2\pi R M^{1,2}}$.
Since small quark masses are naturally realized by the exponential suppression
in the gauge-Higgs unification scenario, i.e. $\frac{m_d^2}{M_W^2} \sim \e^{-2\pi RM^1}$, etc.,
the suppression in the case of LR type diagram is similar to the well-known GIM mechanism and we have called it ``GIM-like" mechanism.
We also have discussed the origin of such qualitative difference of two suppression mechanisms.     

We have shown that although the condition by Glashow-Weinberg \cite{GW} is satisfied
in good approximation for the zero-mode sector of quarks,
we still have FCNC at the tree level because of the presence of new source of flavor violation,
i.e. the non-degenerate bulk masses as a new feature of the gauge-Higgs unification scenario.

The rate of $D^0-\bar D^0$ mixing at the tree level via KK gluon exchange has been calculated.
The obtained result for the mass difference of neutral $D$ meson is suppressed
by the inverse powers of compactification scale $M_{\rm c} = R^{-1}$
(the decoupling effects of heavy non-zero KK gluons).
The contributions from the diagrams without chirality flip (LL and RR) turn out to be dominant.
Then by use of the constraints on the Wilson coefficients
of relevant 4-Fermi effective operators \cite{GGNP}
obtained from the recent measurements of the mass difference by BABAR and Bell experiments
we have obtained the lower bounds on the compactification scale as the prediction of our model.
Though the result depends on a free parameter $\theta'$,
it is of order ${\cal O}({\rm TeV})$ for almost all range of $\theta'$.
The obtained lower bound is much milder
than we naively expect assuming only the decoupling of KK gluons
and should be the reflection of the presence of the suppression mechanism of FCNC,
mentioned above.

In the appendix we re-analyze $K^0$--$\bar K^0$ mixing
including the contributions of LL and RR type diagrams
in addition to the one of LR type diagram calculated in our previous paper \cite{AKLM}.
As is seen there the obtained lower bound on the compactification scale is much more severe
in the case of $K^0$--$\bar K^0$ mixing.
Such difference should be attributed to the difference of the factor of flavor mixing
in the vertices of KK gluons for up-type and down-type quarks.
The origin of such difference should be the difference of up-type and down-type quark masses,
since we easily find that in the imaginative limit
of degenerate up- and down-type quark masses, $a = b = \frac1{\sqrt2}$,
such difference disappears.
We, however, have not completely understood qualitatively
why such difference of the order of magnitude
in the obtained lower bounds for $K^0$--$\bar K^0$ and $D^0$--$\bar D^0$ mixings arises.

Finally we briefly comment on other closely related typical FCNC and CP violating observables in our model: 
i.e. $B^0$--$\bar B^0$ mixing and CP violating parameter $\epsilon$ in the neutral $K$ system. 

Concerning $B^0$--$\bar B^0$ mixing, expanding our model in order to include the third generation is clear necessary. 
A serious issue is how to implement the $t$ quark mass, since the bulk mass is effective only for light quarks. 
It has been pointed out that to introduce 4-th rank symmetric repr. of SU(3), i.e. 15 repr. is necessary to get $m_{t} \simeq 2 M_{W}$ \cite{CCP}. 
Still remaining small gap $m_{t} - 2M_{W}$ is argued to be attributed to the quantum correction. 
Interestingly, in our framework the large gap between $m_{t}$ and $m_{b}$ and small generation mixings between the third generation and lighter generation 
seems to be inevitable consequences. 
Namely, for the third generation the bulk mass is not needed $M_{3} = 0$, as it only works to reduce the quark masses. 
Still the relation $m_{t}^{2} + m_{b}^{2} = (2 M_{W})^{2}$ (at the tree level) implies that in order to guarantee $m_{t} \simeq 2 M_{W}$, 
$m_{b}$ inevitably becomes much smaller than $m_{t}$ in accordance with reality. 
Also to keep $m_{t} \simeq 2 M_{W}$, the mixing between the third generation and lighter generations should be small, 
since such mixing tends to reduce the $t$ quark mass. 
In this way, the large top quark mass seems to lead to the desirable pattern of quark masses and generation mixings. 
Thus in the 0-th order approximation, $U_{3}, \ U_{4}$ in (\ref{Umatrices}) for the three generation model looks like 
\begin{align}
  U_4 
 & 
 = \left[\begin{array}{ccc}
	  \cos\theta' 
	& -\!\sin\theta' 
	& 0 \\[3pt]
	 \sin\theta' 
	& \cos\theta' 
	& 0 \\[3pt]
	 0 
	& 0 
	& 1 
   \end{array}\right]\!\! 
   \left[\begin{array}{ccc}
	a & 0 & 0\\[3pt] 0 & b & 0\\[3pt] 0 & 0 & c
   \end{array}\right], \\[3pt]
  U_3
 & 
  = \left[\begin{array}{ccc}
	  \cos\theta 
	& -\!\sin\theta 
	& 0 \\[3pt]
	  \sin\theta 
	& \cos\theta 
	& 0 \\[3pt]
	  0 
	& 0 
	& 1 
   \end{array}\right]\!\! 
   \left[\begin{array}{ccc}
	\sqrt{1-a^2} & 0 & 0\\[3pt] 0 & \sqrt{1-b^2} & 0\\[3pt] 0 & 0 & \sqrt{1-c^2}
   \end{array}\right]. 
\end{align}
where the parameter $c$ is very close to 1, $c \simeq 0.9983$. 
Thus the third generation is isolated from remaining generations and clearly $B^0$--$\bar B^0$ mixing disappears in this limit. 
This argument implies that even though the $B^0$--$\bar B^0$ mixing  arises after the inclusion of small mixings between the third generation and lighter generations, 
the rate of the FCNC process is suppressed by the small mixings and the lower bound on the compactification scale 
obtained from the comparison of the prediction of our model with the data is expected not to exceed that obtained 
from $D^0$--$\bar D^0$ mixing or $K^0$--$\bar K^0$ mixing discussed in this paper. 

In the analysis of $\epsilon$ in the neutral $K$ system, inclusion of CP violating parameter, so far ignored, is clearly needed. 
In the base where the bulk mass term is diagonalized, only source of the  CP phase will be in $U_{3}$ and $U_{4}$ in (\ref{Umatrices}). 
Interestingly, in our model the CP phase seems to remain even in the two generation scheme. 
Namely the most general form of $U_{3}$ and $U_{4}$ in the two generation scheme is known to be given as 
\begin{align} 
	U_4
  = \!\left[
	\begin{array}{cc}
	 \cos \theta' & -\sin \theta' \\
	 \sin \theta' & \cos \theta' \\
	\end{array}
	\right]\!\!
	\left[
	\begin{array}{cc}
	 a & 0 \\
	 0 & b \\
	\end{array}
	\right], \quad 
	U_3
  = \!\left[
	\begin{array}{cc}
	 \cos \theta & -\sin \theta \\
	 \sin \theta & \cos \theta \\
	\end{array}
	\right]\!\!
 \left[
	\begin{array}{cc}
	1 & 0 \\
	0 & e^{i\delta} \\
	\end{array}
	\right] \!\!
\left[
	\begin{array}{cc}
	 \sqrt{1-a^2} & 0 \\
	 0 & \sqrt{1-b^2} \\
	\end{array}
	\right]. 
\end{align}
where the CP violating phase $\delta$ does not need to appear in $U_{3}$ but may appear in $U_{4}$ if we wish. 
The important observation is that by generalizing the rotation matrices in the left of $U_{3}, \ U_{4}$ 
used in the analysis of $D^0$--$\bar D^0$ and  $K^0$--$\bar K^0$ mixings to unitary matrices, 
we first get 3 phases in each of $U_{3}, \ U_{4}$, but after rephasing of $Q_{3}, \ Q_{6}$ and $Q_{SM}$ there remains only one physical CP violating phase, 
which we wrote $\delta$. Let us note that an overall phase of $Q_{3}, \ Q_{6}$ and $Q_{SM}$ is irrelevant in the reduction of the number of phases. 
Thus it turns out that the KK gluon vertex of the LL type contains the CP violating phase, since it gets contributions from both of $U_{3}$ and $U_{4}$. 
Thus this vertex is expected to give the imaginary part in the amplitude of $K^0$--$\bar K^0$ mixing through KK gluon exchange at the tree level,  
and therefore to the parameter $\epsilon$. 
We do not perform the detailed calculation of $\epsilon$ here, 
but the comparison of our prediction with the data of $\epsilon$ will be another interesting issue to be addressed 
in order to get meaningful constraints on the compactification scale and/or the CP phase $\delta$.     
 
\subsection*{Acknowledgments}

This work was supported in part by the Grant-in-Aid for Scientific Research 
of the Ministry of Education, Science and Culture, No. 21244036.

\appendix

\section{The constraint from $\bs{K^0}$--$\bs{\bar K^0}$ mixing}
\begin{figure}[h]
\[
\begin{array}{ccc}
\includegraphics[bb= 0 0 106 70, scale=1]{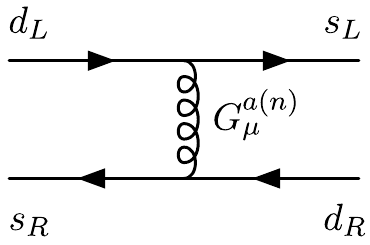} \qquad & \qquad
\includegraphics[bb= 0 0 106 70, scale=1]{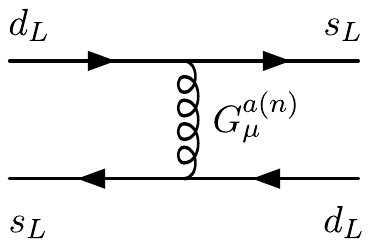} \qquad & \qquad
\includegraphics[bb= 0 0 106 70, scale=1]{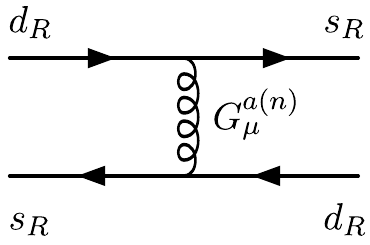}\\[8pt]
\text{(i) LR type} \qquad & \qquad
\text{(ii) LL type} \qquad & \qquad
\text{(iii) RR type}
\end{array}
\]
\caption{The diagrams of $K^0-\bar K^0$ mixing via KK gluon exchange}
\label{fig4}
\end{figure}

We discuss the constraint on the compactification scale $R^{-1}$ 
obtained from the analysis of $K^0$--$\bar K^0$ mixing. 
The contribution of non-zero KK gluon exchange processes is dominant in this case \cite{AKLM}.
In the previous analysis \cite{AKLM}, 
we have considered only the LR type diagram shown in the Fig.\,\ref{fig4}\,(i).
This is because we expected that the contribution of this diagram was dominant
due to the chiral enhancement factor $\frac{m_K}{m_d+m_s}$ in the hadronic matrix element 
under our assumption that the mode sums 
$S_\text{KK}^{LL}\big(\!=S_\text{KK}^{RR}\big)$ and $S_\text{KK}^{LR}$
given in (\ref{KKbarcontri})
are more or less of the same order of magnitude.
However, it turns out that we have to take into account the other contributions 
from the LL and the RR type diagrams as well, since the above naive assumption on 
the mode sum is not correct as the matter of fact, $S_\text{KK}^{LL} \gg S_\text{KK}^{LR}$.  
Therefore, we re-analyze $K^0-\bar{K}^0$ mixing in this appendix.

\begin{figure}[t]
\begin{equation}
\begin{array}{cc}
   \includegraphics[bb=0 0 600 480, scale=0.4]{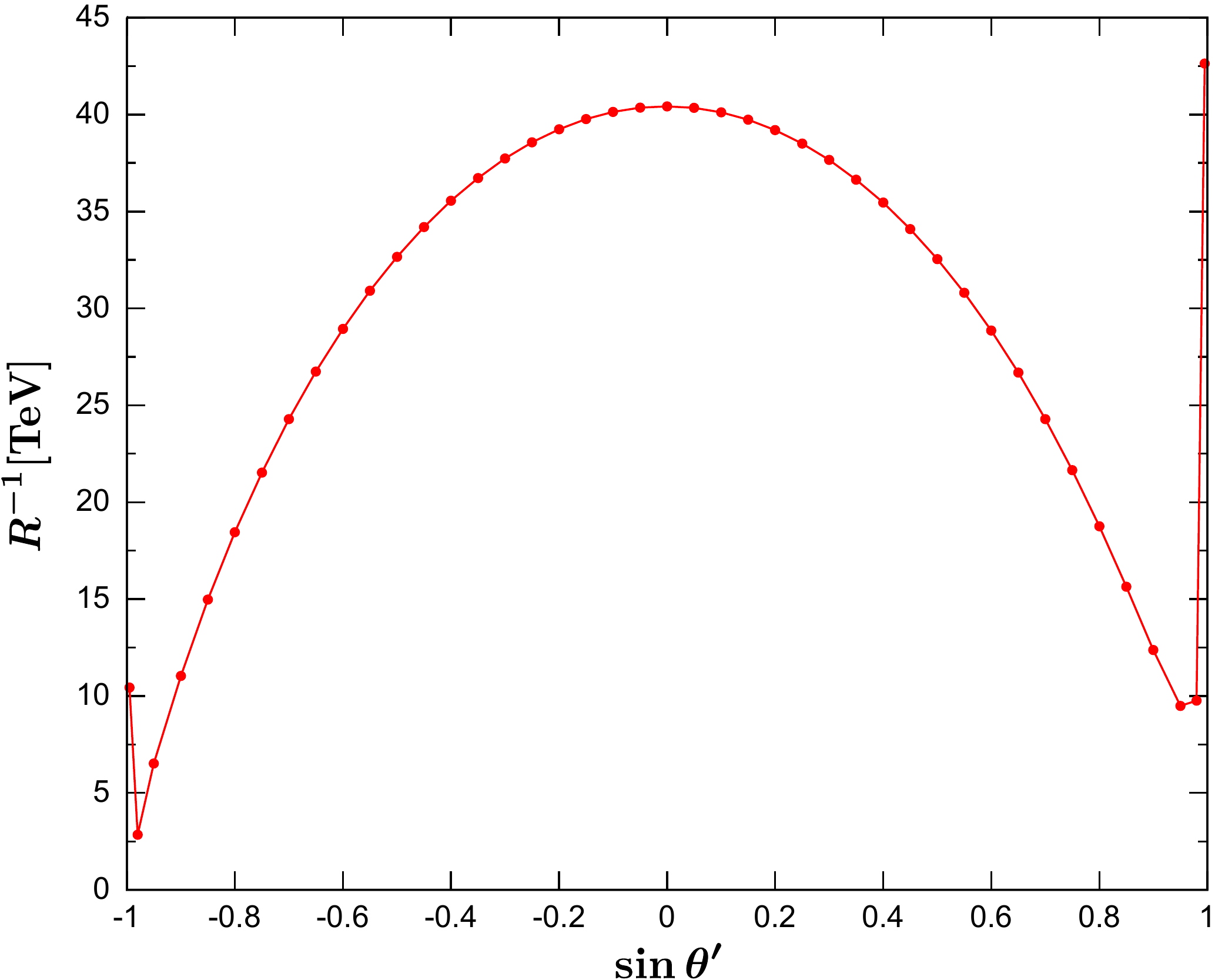}
\end{array}\notag
\end{equation}
\caption{Lower bounds on $R^{-1}$  obtained from the data on $K^0$--$\bar K^0$ mixing.}
\label{fig5}
\end{figure}

The contributions to $K_L-K_S$ mass difference $\Delta m_K(\text{KK})$
from each type of diagrams shown in Fig.\,\ref{fig4} are given as follows, respectively;
\begin{align}
	\Delta m_K^{LR}(\text{KK})
 &= 2\pi\alpha_{\rm s}R^2\!
	\left(B_4-\frac{B_5}9\right)\!\!
	\left(\frac{m_K}{m_d+m_s}\right)^{\!\!2}\!
	f_K^2 m_K
	\sin2\theta_{dR}(\alpha_d+\alpha'_d) S_\text{KK}^{LR}\ ,\notag\\
	\label{KKbarcontri}
	\Delta m_K^{LL}(\text{KK})
 &= -\frac49\pi B_1\alpha_{\rm s}R^2f_K^2 m_K
	(\alpha_d+\alpha'_d)^2 S_\text{KK}^{LL}\ ,\\[3pt]
	\Delta m_K^{RR}(\text{KK})
 &= -\frac49\pi B_1\alpha_{\rm s}R^2f_K^2 m_K
	\sin^2\!2\theta_{dR}S_\text{KK}^{LL}\ .\notag
\end{align}
where 
\begin{align}
	\alpha_d
 \equiv& 
	-\!(1-a^2)\sin2\theta_{dL} \cos^2\!\theta
	+(1-b^2)\sin2\theta_{dL} \sin^2\!\theta
	-\!\sqrt{(1-a^2)(1-b^2)}\cos2\theta_{dL} \sin2\theta\ ,\notag\\[2pt]
	\alpha'_d
 \equiv& \label{alphadprime}
	-a^2\sin2\theta_{dL} \cos^2\!\theta'+b^2\sin2\theta_{dL} \sin^2\!\theta'
	-ab\cos2\theta_{dL} \sin2\theta'. 
\end{align}
$\theta_{dR}$ is an angle in the rotation matrix $V_{dR}$
to diagonalize $I_{RL}^{(00)}U_3 U_3^\dag I_{RL}^{(00)}$:
\begin{equation}
   \tan2\theta_{dR} 
 = \frac{2\big(b^2-a^2\big)cd\sin\theta\cos\theta}
		 {(1-a^2)c^2-(1-b^2)d^2+(a^2-b^2)(c^2+d^2)\sin^2\!\theta}.
\end{equation}
The bag parameters are calculated by lattice simulation 
as $B_1=0.57$, $B_4=0.81$ and $B_5=0.56$ \cite{BBDGW}. 
$f_K(\simeq 1.23f_\pi), m_K$ is the kaon decay constant and the kaon mass, respectively.
The constant $\alpha_{\s}$ is estimated to be $\alpha_\s(\mu_K) \approx 0.268$ for $\mu_K = 2.0$\,GeV \cite{BBDGW}.
Combining these results, we obtain
\begin{align}
	\Delta m_K(\text{KK})
 \sim&
	-\!1.61\times10^2
	\cdot(Rf_\pi)^2\notag\\*
 &\times\!
	\bigg[
	\Big\{(\alpha_d+\alpha'_d)^2 + \sin^2\!2\theta_{dR}\Big\}
	S_\text{KK}^{LL}
	-65.0
	\cdot
	\sin2\theta_{dR}(\alpha_d+\alpha'_d)
	S_\text{KK}^{LR}
	\bigg]\,[{\rm MeV}]
\end{align}
The room for the \lq\lq New Physics\rq\rq contribution $\Delta m_K(\text{NP})$ is basically given
by the difference between the experimental data
and the standard model prediction \cite{IL}, \cite{BS}. Though the short-distance contribution due to the box diagram 
in the standard model \cite{IL} is reliably calculated the long-distance contribution has uncertainty. Thus here we take an attitude that $\Delta m_K(\text{NP})$ can be as large as the experimental value: 
\begin{align}
	\big|\Delta m_K(\text{NP})\big|
  &< \Delta m_K ({\rm Exp})
  =  3.48 \times 10^{-12}~[{\rm MeV}]
\end{align}
Identifying $\Delta m_K(\text{NP})$ with our result $\Delta m_K(\text{KK})$ 
we obtain a lower bound for the compactification scale:  
\begin{align}
	R^{-1}
 &\gtrsim
	6.32\times10^2
	\sqrt{
	\Big\{(\alpha_d+\alpha'_d)^2 +\sin^2\!2\theta_{dR}\Big\}S_\text{KK}^{LL}
	-65.0\cdot\sin2\theta_{dR}(\alpha_d+\alpha'_d)S_\text{KK}^{LR}
	}~[{\rm TeV}]\ .
	\label{KKbound}
\end{align}
The obtained numerical result is given in Fig.\,\ref{fig5}
where the first term inside the square root in (\ref{KKbound}) is dominant. 
\noindent
The lower bound on the compactification scale $R^{-1}$ ranges
from 2.8\,TeV to 43\,TeV depending on the value of $\sin\theta'$.


\end{document}